\documentclass{article}

\usepackage{PRIMEarxiv}

\usepackage[utf8]{inputenc} % allow utf-8 input
\usepackage[T1]{fontenc}    % use 8-bit T1 fonts
\usepackage{hyperref}       % hyperlinks
\usepackage{url}            % simple URL typesetting
\usepackage{booktabs}       % professional-quality tables
\usepackage{amsfonts}       % blackboard math symbols
\usepackage{nicefrac}       % compact symbols for 1/2, etc.
\usepackage{microtype}      % microtypography
\usepackage{fancyhdr}       % header
\usepackage{graphicx}       % graphics
\usepackage{tikz}
\usepackage{array}
\usepackage{tabularx}
\usepackage{amsmath}
\usepackage{dsfont}         % For \mathds
\usepackage{multicol}
\usepackage{circuitikz}
\usepackage{subcaption}
\usetikzlibrary{positioning, arrows.meta, calc, shapes}
\usepackage{xcolor}

\usepackage{amssymb}

\title{A Survey on the Applications of Zero-Knowledge Proofs}

\author{
  Ryan Lavin, Xuekai Liu, Hardhik Mohanty\thanks{Corresponding author: \texttt{hmohanty@usc.edu}}, Logan Norman, Giovanni Zaarour, Bhaskar Krishnamachari \\
  Viterbi School of Engineering \\
  University of Southern California, Los Angeles, California, USA \\
  \texttt{\{rlavin, xuekaili, hmohanty, fnorman, gzaarour, bkrishna\}@usc.edu}
}

\begin{document}
\maketitle

\begin{abstract}
Zero-knowledge proofs (ZKPs) enable computational integrity and privacy by allowing one party to prove the truth of a statement without revealing underlying data. Compared with alternatives such as homomorphic encryption and secure multiparty computation, ZKPs offer distinct advantages in universality and minimal trust assumptions, with applications spanning blockchain systems and confidential verification of computational tasks. This survey provides a technical overview of ZKPs with a focus on an increasingly relevant subset called zkSNARKs. Unlike prior surveys emphasizing algorithmic and theoretical aspects, we take a broader view of practical deployments and recent use cases across multiple domains including blockchain privacy, scaling, storage, and interoperability, as well as non-blockchain applications such as voting, authentication, timelocks, and machine learning. To support consistent comparison, we provide (i) a taxonomy of application areas, (ii) evaluation criteria including proof size, prover and verifier time, memory, and setup assumptions, and (iii) comparative tables summarizing key tradeoffs and representative systems. The survey also covers supporting infrastructure, including zero-knowledge virtual machines, domain-specific languages, libraries, and frameworks. While emphasizing zkSNARKs for their prevalence in deployed systems, we compare them with zkSTARKs and Bulletproofs to clarify transparency and performance tradeoffs. We conclude with future research and application directions.
\end{abstract}

\keywords{zero-knowledge proofs \and zkSNARKs \and zkSTARKs \and bulletproofs \and blockchain \and cryptography \and privacy}

\section{Introduction}

Zero-knowledge proofs (ZKPs) are a set of cryptographic methods allowing one party to prove the validity of a claim to another without disclosing any of the claim's underlying details. The seminal work by Goldwasser, Micali, and Rackoff \cite{goldwasser1985knowledge} laid the groundwork for ZKPs in the 1980s. It introduced the principle of knowledge complexity as a metric for quantifying the information transferred from the prover to the verifier. Subsequently, Goldreich and others \cite{goldreich1991proofs, ben1990everything} published several works until the 1990s that expanded the scope of ZKPs to a broader range of computational problems. They demonstrated their applicability to NP-complete problems under specific cryptographic assumptions. Another significant advancement in the development of ZKPs came with the introduction of succinctness in \cite{BitanskySNARK11, groth10} following the design paradigm of Kilian's seminal paper \cite{kilian1992note}. The succinct and non-interactive properties of zkSNARKs greatly enhance their practical applicability in blockchain and non-blockchain systems. These properties make ZKPs applicable to digital systems that require both security and privacy \cite{zhang2019security}. This survey is written for three primary audiences: (i) Systems and security researchers can use it to understand the application landscape and recurring engineering bottlenecks that shape real deployments, (ii) Blockchain engineers can use it to navigate how ZKPs change system design choices around scalability, privacy, and cost, (iii) Students and applied cryptography practitioners can use it as a guided map from core components to modern toolchains and application patterns.

The advent of ZKPs has represented a significant leap forward in the convergence of privacy and verifiability, which has attracted significant interest from both the cryptographic community and industry. It introduced a unique approach in the field of cryptography, differentiating itself from other privacy-preserving computation methods for distributed systems, such as homomorphic encryption and secure multiparty computation. While these methods are also being actively developed and advanced, each serves a specific purpose in information verification and privacy preservation. Homomorphic encryption enables computations on encrypted data without needing to decrypt, thereby preserving confidentiality while allowing the derivation of useful insights \cite{rivest1978data}. Secure multiparty computation enables trustless collaborations and allows parties to jointly compute a function over their inputs while keeping those inputs private \cite{yao1982protocols}. ZKPs offer a distinct tradeoff profile, often enabling succinct public verification and privacy under assumptions that vary by proof family, by proving the truth of a statement without disclosing any other information.

In modern digital systems, a tradeoff exists between openness and privacy. Blockchains, for example, prioritize transparency to ensure trust and prevent fraud, with every transaction openly verifiable. However, this transparency can compromise privacy \cite{o2019design}. Despite the pseudonymous nature of blockchain transactions, advanced analytics can de-anonymize users by correlating on-chain and off-chain data. Such exposure can reveal a user's entire transaction history, leading to privacy breaches and potential targeted threats. As digital infrastructures become increasingly complex, striking the right balance between transparency for security and preserving user privacy becomes a vital research challenge. In this context, ZKPs emerge as a robust solution addressing the challenges posed by such tradeoffs in digital systems.

ZKPs enable users to verify private data such as bank balances or credit scores without revealing specifics. Furthermore, they can ensure anonymity in authorization processes by allowing access to restricted areas or proving regional origin without disclosing detailed credentials. In the financial domain, ZKPs can allow for identity-free payments and tax submissions without revealing exact earnings. Additionally, they facilitate trustless outsourcing by letting organizations validate results without redoing the entire computation. Furthermore, ZKPs can modify blockchain operations by shifting computation off-chain while preserving network-wide verification. Overall, ZKPs exemplify the merging of verification with privacy across diverse applications \cite{petkus2019and}.

The practical applications of ZKPs rest on verifiable computation: the ability to prove that a computation was performed correctly without revealing the inputs or the computation process. Verifiable computation enables the two main value propositions of ZKPs -- succinctness and privacy. In the remainder of this survey, we will see how each of these two values contributes to the practical applications of ZKPs.
Succinctness in ZKPs allows for the quick verification of the correctness of a computation without the extensive resources typically required for direct computation execution. This matters especially in contexts where computational resources are limited or expensive, such as blockchain networks. For instance, ZKPs enable the validation of transaction blocks or smart contract executions without necessitating each node to replay the entire sequence of transactions or computations. This aspect dramatically reduces the computational load on the network, enhancing scalability and throughput.
Privacy, the second value proposition, follows from the ability of ZKPs to prove correctness without revealing the underlying data. This property is especially useful when sensitive or confidential data is involved, enabling applications such as private voting systems and confidential financial transactions in which process integrity is maintained without exposing participants' information.

While zkSNARKs are emphasized because they are widely used in many deployed systems, especially where small proofs and cheap verification are important, this survey covers ZK applications broadly and compares major proof families, including zkSNARKs, zkSTARKs, and Bulletproofs, to explain when different assumptions and performance profiles are preferred in practice. Throughout the survey, we use a consistent evaluation lens to compare systems and application designs, focusing on proof size, prover time, verifier time, peak memory, setup assumptions (trusted versus transparent), and, for blockchain deployments, on-chain verification cost and data availability cost.

In the ZKP literature, there exist two key survey articles relevant to our work. The first survey by Morais \textit{et al.} \cite{morais2019survey} dives deep into Zero-Knowledge Range Proofs (ZKRP) with a particular emphasis on the algorithmic details required for the implementation of ZKRPs. Furthermore, it defines the fundamental theoretical background behind ZKRPs and describes possible generic and blockchain-specific use cases of ZKRPs. In contrast, our survey covers a broader range of application categories and use cases. Further, we provide an overview of the underlying theoretical components, tools, and infrastructure needed for ZKP development. Additionally, we enlist and discuss currently deployed and in-deployment application software that fall under different blockchain and non-blockchain categories. The second survey by Sun \textit{et al.} \cite{sun2021survey} takes a narrower approach, concentrating on the intersection of ZKPs with blockchain technology. It outlines the security challenges within blockchain's public and transparent nature and discusses how ZKPs can mitigate risks of private data exposure. Our work differentiates itself from these prior surveys by providing a more comprehensive practical application-oriented overview that covers both blockchain and non-blockchain applications of ZKPs, complemented by an analysis of the ZKP infrastructure being developed, including zkVMs, DSLs, and supportive technologies to guide developers to essential tools for building ZK-enabled applications efficiently.

This survey provides a comprehensive overview for practitioners and researchers, covering the practical applications and use cases of ZKPs. It describes best practices, challenges, and the impact of ZKPs across application fields. For practitioners, it offers guidance on incorporating ZKPs into real systems. For researchers, it summarizes the current state of ZKP deployment and identifies open problems. By connecting the theoretical cryptographic foundations to practical application domains, the survey aims to help professionals across disciplines apply ZKPs effectively.

We organize the remainder of this article to methodically illustrate the components, toolchains, and applications of ZKPs using a consistent evaluation lens. Section \ref{sec:components} defines core primitives, terminology, and the foundational components and requisite properties used throughout the survey. Section \ref{sec:infra} describes the infrastructure underpinning ZKPs, encompassing virtual machines, domain-specific languages, and pertinent libraries. Next, Section \ref{sec:applications} reviews blockchain applications and compares system designs using the evaluation criteria described above. Section \ref{sec:nonblockchain} extends the discussion to non-blockchain application domains, detailing ZKPs' utility in contexts such as machine learning and digital identity verification. Finally, Section \ref{sec:conclusion} discusses open challenges and future directions, then wraps up with a summary and reflective analysis. The overall survey structure is depicted in Figure~\ref{fig:zkp_survey_structure}.

\begin{figure}[!ht]
\centering
\resizebox{0.8\linewidth}{!}{%
\begin{tikzpicture}[
  font=\large,
  >={Stealth},
  container/.style={
    draw, rounded corners=18pt, thick,
    fill=gray!6, inner sep=14pt
  },
  title/.style={font=\Large, align=center},
  box/.style={
    draw, rounded corners=12pt, thick,
    minimum width=5.0cm, minimum height=1.05cm,
    align=center, fill=white
  },
  cat/.style={
    draw, rounded corners=12pt, thick,
    minimum width=7.2cm, minimum height=1.05cm,
    align=center, fill=white
  },
  sub/.style={
    draw, rounded corners=12pt, thick,
    minimum width=3.8cm, minimum height=0.95cm,
    align=center, fill=white
  },
  bigarr/.style={->, line width=2.6pt}
]

% Centers
\def\DEVY{8.0}
\def\APPY{0.1}

% =======================
% Top container: Dev tools
% =======================
\node[container, minimum width=17.5cm, minimum height=4.6cm] (DEV) at (0,\DEVY) {};
\node[title] at ($(DEV.north)+(0,-0.55)$) {ZK Developer Tools};

\node[box] (zkvm)  at ($(DEV.center)+(-6.0,0.35)$)  {ZK Virtual\\Machines};
\node[box] (zkdsl) at ($(DEV.center)+( 0.0,0.35)$)  {ZK Domain-Specific\\Languages};
\node[box] (zklib) at ($(DEV.center)+( 6.0,0.35)$)  {ZK Libraries \&\\Frameworks};
\node[box] (zkhw)  at ($(DEV.center)+( 0.0,-1.25)$) {ZK Hardware\\Acceleration};

% =========================
% Bottom container: Apps
% =========================
\node[container, minimum width=17.5cm, minimum height=7.4cm] (APP) at (0,\APPY) {};
\node[title] at ($(APP.north)+(0,-0.55)$) {ZKP Applications};

% Categories
\node[cat] (bc)  at ($(APP.center)+(-4.2,2.2)$)  {Blockchain Applications};
\node[cat] (nbc) at ($(APP.center)+( 4.6,2.2)$)  {Non-Blockchain Applications};

% Blockchain subcategories
\node[sub] (poi_b)   at ($(APP.center)+(-6.4,0.9)$)   {Proof of\\Identity};
\node[sub] (l1)      at ($(APP.center)+(-2.0,0.9)$)   {Layer 1\\Blockchains};

\node[sub] (sce)     at ($(APP.center)+(-6.4,-0.4)$)  {Supply Chain /\\Enterprise};
\node[sub] (l2)      at ($(APP.center)+(-2.0,-0.4)$)  {Layer 2\\Scaling};

\node[sub] (interop) at ($(APP.center)+(-6.4,-1.7)$)  {Interoperability};
\node[sub] (txp)     at ($(APP.center)+(-2.0,-1.7)$)  {Transaction\\Privacy};

\node[sub] (stor)    at ($(APP.center)+(-6.4,-3.0)$)  {Storage};
\node[sub] (por)     at ($(APP.center)+(-2.0,-3.0)$)  {Proof of\\Reserves};

% Non-blockchain subcategories
\node[sub] (poi_nb) at ($(APP.center)+(4.6,0.7)$)   {Proof of Identity};
\node[sub] (ml)     at ($(APP.center)+(4.6,-0.6)$)  {ZK ML/AI};
\node[sub] (other)  at ($(APP.center)+(4.6,-1.9)$)  {Other Applications};

% =========================
% Upward arrow: Apps -> Tools
% =========================
\draw[bigarr]
  ($(APP.north)+(0,0.15)$) -- ($(DEV.south)+(0,-0.15)$)
  node[midway, fill=white, inner sep=2pt, font=\small]
  {Implemented using};

\end{tikzpicture}%
}
\caption{ZKP Applications \& Developer Tools (Survey Structure).}
\label{fig:zkp_survey_structure}
\end{figure}

\section{Foundations of Zero-Knowledge Proofs}
\label{sec:components}

Zero-knowledge proofs (ZKPs) are cryptographic protocols that enable one party, known as the prover, to convince another party, the verifier, that a statement is true without revealing any information beyond the validity of the statement itself \cite{goldwasser1985knowledge}. In this context, a statement is a claim that can be efficiently verified, such as knowledge of factors of a large integer, or correct execution of a program under a specified instruction set. This section clarifies the relationship between the broad class of ZKPs and two widely used refinements: SNARKs and zkSNARKs \footnote{Our overview is necessarily brief. For a more detailed discussion, we refer the reader to explainers such as~\cite{petkus2019and}.}.

\subsection{Definitions and Core Properties}
\label{sec:defs_properties}

To avoid ambiguity, we use the following layered terminology throughout the survey.

\begin{itemize}
    \item \textbf{ZKP:} A zero-knowledge proof is a protocol for proving that a statement is true without revealing the witness. ZKPs may be interactive or non-interactive, and they may be succinct or non-succinct depending on the construction.
    \item \textbf{SNARK:} A SNARK is a \textit{succinct} and \textit{non-interactive} \textit{argument of knowledge}. Importantly, being a SNARK does not automatically imply the zero-knowledge property. A SNARK may or may not be instantiated with a zero-knowledge guarantee, depending on how it is constructed.
    \item \textbf{zkSNARK:} A zkSNARK is a SNARK that additionally satisfies the zero-knowledge property, meaning that the proof reveals nothing about the witness beyond the fact that the statement is true.
\end{itemize}

The term \textit{argument} distinguishes computational soundness from information-theoretic soundness. In a proof system with information-theoretic soundness, no prover (even unbounded) can convince the verifier of a false statement. In an argument system, soundness is computational: it holds against any efficient (polynomial-time) prover under standard cryptographic assumptions. SNARKs are arguments because their security relies on computational assumptions.

We summarize the most commonly referenced properties below. When a construction is specifically a zkSNARK, it includes the zero-knowledge property in addition to succinctness, non-interactivity, and knowledge soundness.

\begin{enumerate}
    \item \textbf{Completeness:} If the statement is true and both parties follow the protocol, the verifier accepts a valid proof.
    \item \textbf{Soundness (computational):} If the statement is false, no efficient prover can convince the verifier to accept, except with negligible probability.
    \item \textbf{Knowledge soundness:} If a prover convinces the verifier, then the prover knows a valid witness for the statement (formalized via an extractor).
    \item \textbf{Zero knowledge (when present):} The proof leaks no information about the witness beyond statement validity. Depending on the construction, this can be computational, statistical, or perfect.
\end{enumerate}

Many deployed zkSNARK systems rely on public parameters generated in a setup phase. A \textit{common reference string} (CRS) refers to these public parameters that are shared by prover and verifier. In some constructions, the CRS is generated together with secret randomness (often called a trapdoor). If the trapdoor is ever learned by an adversary, it may enable forging proofs or breaking soundness, depending on the specific scheme. This is why CRS generation is often done via a multi-party ceremony designed so that as long as one participant is honest and discards their secret randomness, the trapdoor remains unknown. In contrast, \textit{transparent} setups aim to avoid secret trapdoors by deriving public parameters from public randomness and standard assumptions, so that the setup does not rely on the secrecy of any toxic waste. Ongoing standardization efforts, such as those coordinated by ZKProof.org~\cite{zkprooforg}, are formalizing setup categories and interoperability requirements across proof systems.

\textbf{Fiat--Shamir versus setup:}
Non-interactivity and setup are related but distinct concepts. The \textit{Fiat--Shamir heuristic} \cite{fiat1986prove} is a transform that turns certain interactive proof protocols into non-interactive ones by replacing the verifier's random challenges with outputs of a hash function (typically modeled as a random oracle). This addresses interactivity. Separately, \textit{trusted setup} versus \textit{transparent setup} concerns how public parameters are generated and whether they depend on secret trapdoor material. A system can be non-interactive while still requiring a trusted setup, and it can also be non-interactive with a transparent setup, depending on the proof family.

\begin{table}[t]
\centering
\caption{High-level tendencies of common proof-system families used in practice. These entries are broad tendencies rather than universal properties.}
\label{tab:properties_matrix}
\begin{tabular}{lccccc}
\hline
\textbf{Category} & \textbf{Zero Knowledge} & \textbf{Succinct} & \textbf{Interactive} & \textbf{Setup} & \textbf{Post-quantum} \\
\hline
ZKP (general)     & yes & optional & optional & optional & depends \\
SNARK             & optional & yes      & no       & trusted  & no \\
zkSNARK           & yes      & yes      & no       & trusted & no \\
zkSTARK           & yes      & yes      & no       & transparent & yes \\
Bulletproofs      & yes      & no & no & transparent & typically no \\
\hline
\end{tabular}
\end{table}

\subsection{Lifecycle of a SNARK: From Python to Polynomials}
\label{sec:SNARKlifecycle}

\begin{figure}[h!]
    \centering
    \includegraphics[width=0.78\linewidth]{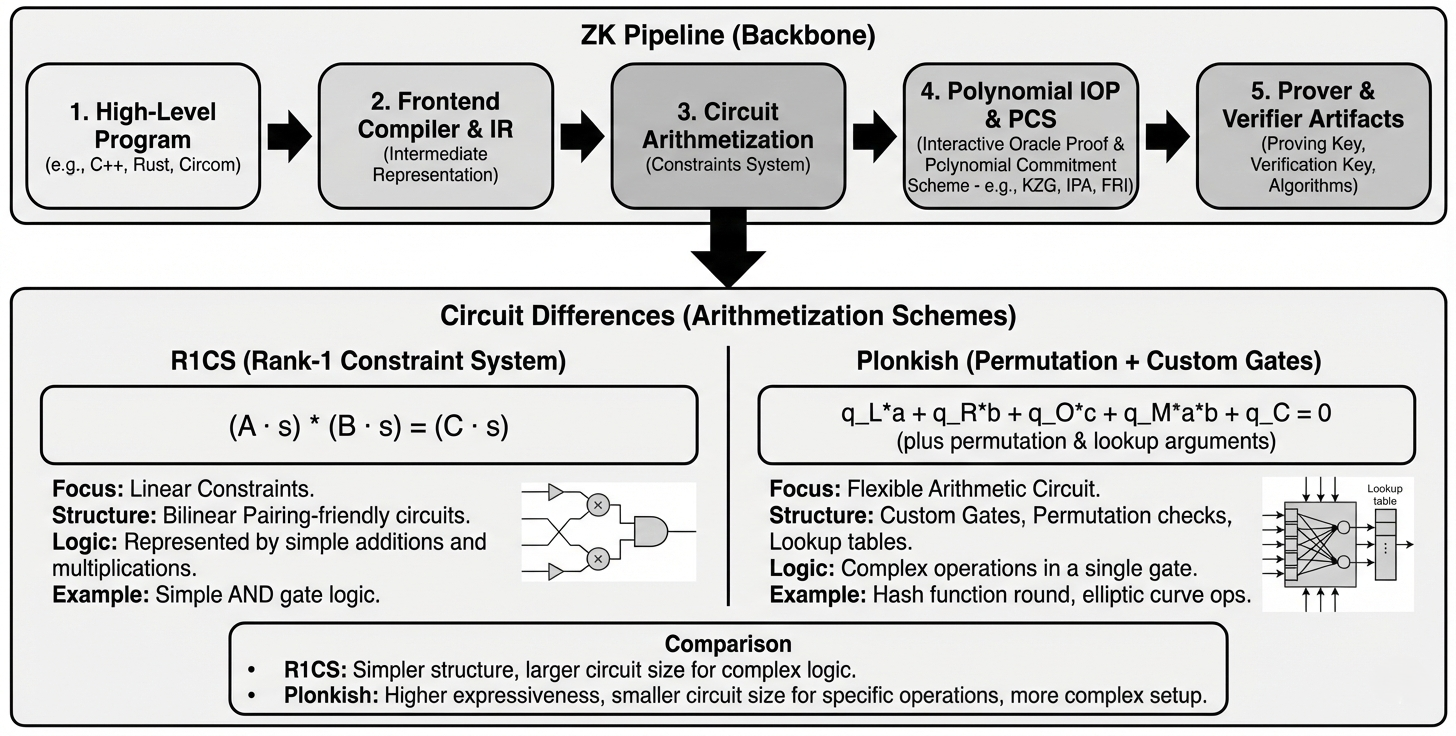}
    \caption{ZKP Pipeline:  The frontend compiles programs into a constraint system; the backend combines a polynomial IOP with a polynomial commitment scheme (PCS) producing the succinct proof and verification key.}
    \label{fig:structure}
\end{figure}

\subsubsection{Frontends: From High-level code to circuits}

The Python function\footnote{\textbf{A practical caveat:} While the transition from Python code to an arithmetic circuit illustrates the conceptual process underlying SNARK construction, directly translating a general-purpose language into constraints is often inefficient. In practice, developers commonly use zero-knowledge domain-specific languages (zkDSLs) and compilers designed for constraint generation. These are explored in Section \ref{sec:DSLs}.} (Figure~\ref{fig:structure}) calculates the evaluation of a three-variate polynomial \( f(a, b, c) = a^2 b^2 + b^2 c^2 \). Next, we represent this function as an arithmetic circuit. Arithmetic circuits decompose complex computations into simple arithmetic operations (such as addition and multiplication). This transformation, together with later steps described below, enables the privacy and succinctness properties that make modern SNARK systems practical.

In cryptographic circuits, the elements \( a \), \( b \), and \( c \) are typically treated as elements of a finite field \(\mathds{F}_p\), that is, integers modulo a large prime \(p\). This ensures that all operations are performed within a closed set \(\{0, \ldots, p-1\}\), enabling algebraic techniques used by modern proving systems.

\subsubsection{Arithmetization: From circuits to matrices}
\label{sec:arithmetization}

\begin{table}[t]
\centering
\caption{High-level comparison of arithmetization strategies.}
\label{tab:arithmetization}
\resizebox{\columnwidth}{!}{%
\footnotesize
\begin{tabular}{l p{2.7cm} p{2.7cm} p{2.7cm} p{2.7cm}}
\hline
\textbf{Property} & \textbf{R1CS} & \textbf{Plonkish} & \textbf{AIR} & \textbf{CCS (folding)} \\
\hline
Constraint style &
Bilinear constraints on linear combos &
Custom gates + permutation/copy constraints &
Polynomial constraints over an execution trace &
Generalized constraints designed for folding/IVC \\
Strengths &
Simple, mature tooling &
Flexible arith., efficient lookups &
Transparent (no trusted setup), good for large traces &
Enables efficient folding / incremental verification \\
Typical prover cost driver &
MSMs (and pairings in some systems) &
FFT/NTT + MSM &
Hashing + large-trace memory/IO &
Backend-dependent \\
Lookups &
Usually emulated (or via extra gadgets) &
Native (Plookup/LogUp/Lasso, etc.) &
Via lookup arguments (e.g., LogUp) &
Via composition within the folded system \\
Recursion / IVC &
Often pairing-cycle-based in pairing SNARKs &
Accumulation / recursion-friendly designs &
FRI-style recursion/composition &
Folding (Nova, HyperNova, etc.) \\
Examples &
Groth16, Marlin &
Plonk, Halo2 &
STARKs, Cairo &
Nova, HyperNova \\
\hline
\end{tabular}%
}
\end{table}

Arithmetization is the process of encoding the wiring and gate operations of a circuit or model of computation into an algebraic representation suitable for proof generation. For brevity, we focus on Rank-1 Constraint Systems (R1CS), a common arithmetization strategy for many zkSNARK constructions. Alternative arithmetizations, including Plonkish arithmetization, algebraic intermediate representations (AIR), and circuit constraint systems (CCS) (see Table~\ref{tab:arithmetization}), have gained popularity in recent years \cite{ben2013snarks, settyThalerWahbyCCS23}.

An R1CS represents a computation as constraints of the form
\[
(\mathbf{A}\cdot \mathbf{w}) \circ (\mathbf{B}\cdot \mathbf{w}) = (\mathbf{C}\cdot \mathbf{w}),
\]
where \(\mathbf{A}\), \(\mathbf{B}\), and \(\mathbf{C}\) are matrices, \(\mathbf{w}\) is a vector of variables corresponding to the execution transcript (the witness), and \(\circ\) denotes the Hadamard (element-wise) product.

Each gate in the circuit corresponds to one constraint row. For our example, six constraints encode the three squarings ($a \cdot a = a^2$, $b \cdot b = b^2$, $c \cdot c = c^2$), the two cross-multiplications ($a^2 \cdot b^2$, $b^2 \cdot c^2$), and the final addition.

To encode these constraints into matrices \( \mathbf{A} \), \( \mathbf{B} \), and \( \mathbf{C} \), we introduce variables: \( w_1 = a \), \( w_2 = b \), \( w_3 = c \), \( w_4 = a^2 \), \( w_5 = b^2 \), \( w_6 = c^2 \), \( w_7 = a^2b^2 \), \( w_8 = b^2c^2 \), and \( w_9 = \text{result} \). A compact witness vector representation is then
\[
\mathbf{w} = [1, a, b, c, a^2, b^2, c^2, a^2b^2, b^2c^2, \text{result} ].
\]

The witness \(\mathbf{w}\) captures a valid execution transcript of the circuit by assigning values to all intermediate variables so that every constraint is satisfied. If the witness were revealed, it would leak inputs and intermediate values. Up to this point, however, there is still no way to prove correctness succinctly and publicly without allowing others to re-run the computation. The next step converts the R1CS into a polynomial form that enables succinct public verification while preserving privacy in zkSNARK settings.

\subsubsection{Backends: From matrices to polynomials}

The backend phase transforms the R1CS constraints into polynomial equations, commonly instantiated as a Quadratic Arithmetic Program (QAP) \cite{GGPR13}. This transformation enables succinct verification by reducing many constraint checks into a small number of algebraic relations.

For matrices \( \mathbf{A}, \mathbf{B}, \mathbf{C} \) of size \( m \times n \), where \( m \) is the number of constraints and \( n \) is the number of variables, we define polynomials \( \{A_i(x)\}_{i=1}^m \), \( \{B_i(x)\}_{i=1}^m \), and \( \{C_i(x)\}_{i=1}^m \) corresponding to the rows of \( \mathbf{A}, \mathbf{B}, \mathbf{C} \), respectively. The QAP condition can be expressed as:
\[
A(x)\cdot B(x) - C(x) = h(x)\cdot Z(x),
\]
where \(h(x)\) encodes the satisfaction of the constraints, and \(Z(x)\) is a fixed target polynomial that vanishes on the constraint index set.

A prover can produce a compact proof that such an \(h(x)\) exists, which implies that the prover knows a witness \(\mathbf{w}\) satisfying the original R1CS. A verifier can then check the required polynomial relations efficiently, without re-executing the full computation, yielding a substantial speedup in verification time.

\subsection{Conjoining Information Theory and Cryptography}

Interactive proofs (IPs), when considered without cryptographic assumptions, can provide information-theoretic soundness. Zero knowledge means that whatever the verifier learns from the interaction could also have been produced by the verifier on its own, without access to the witness, so the protocol leaks nothing beyond the fact that the statement is true. Modern SNARK designs integrate algebraic proof techniques with cryptographic components to achieve succinctness and public verifiability. A common modern design pattern is to combine a polynomial interactive-oracle proof (PIOP/IOP) that reduces computation to a few algebraic checks with a polynomial commitment scheme (PCS), and then apply Fiat--Shamir to remove interaction in the random-oracle model. A common cryptographic component is a polynomial commitment scheme (PCS) \cite{kate2010constant}, which allows the prover to commit to polynomials and later open evaluations in a way that is binding and (often) hiding. This enables efficient verification while preserving soundness under computational assumptions.

Setup assumptions depend on the proof family. In many pairing-based zkSNARK systems, public parameters are generated via a CRS that may depend on secret trapdoor randomness, motivating multi-party ceremonies to reduce trust in any single participant. In transparent families, public parameters can be derived from public randomness, avoiding secret trapdoors.

Finally, non-interactivity is typically achieved either by designs that are non-interactive by construction or by applying the Fiat--Shamir heuristic \cite{fiat1986prove} to suitable interactive protocols, replacing verifier challenges with hash-derived values under a random-oracle-style assumption. This step enables public verifiability and makes the resulting proofs practical for deployment in broad applications.

\section{ZKP Software Tools and Platforms} \label{sec:infra}
We organize this section using a simple toolchain taxonomy that will be reused in later application comparisons: (i) \emph{zkVMs} provide a general-purpose execution layer that produces proofs of program execution, (ii) \emph{zkDSLs and circuit languages} compile high-level intent into circuits or constraints, (iii) \emph{libraries and frameworks} provide reusable gadgets, curve and field arithmetic, and proof-system implementations, (iv) \emph{proving backends} implement the concrete proof-system algorithms and polynomial commitment schemes, and (v) \emph{hardware accelerators} (GPU/FPGA/ASIC) target the dominant kernels such as MSM, FFT/NTT, and hashing.

\subsection{Zero-Knowledge Virtual Machines}

\subsubsection{Motivation and Definition}

zkVM is a virtual machine (VM) designed to generate an efficiently verifiable zero-knowledge proof for executing an arbitrary computation while preserving the privacy of the program and its data in zero knowledge. The zkVM acts as a programmable ZK circuit that implements a VM, receives public and private inputs, and creates a certificate of valid execution. Similar to a generalized virtual machine, a zkVM maintains its own virtualized components, such as memory management, instruction scheduling, error handling, and others, using a privacy-focused approach with the help of a ZKP (see Figure~\ref{fig:zkvm_arch}). zkVMs simplify ZKP creation. Traditionally, developers needed cryptography expertise, circuit representation skills, backend setup for each proof, and significant computing power. zkVMs abstract these complexities, streamlining the process for developers.

\subsubsection{Methodology}

In Figure~\ref{fig:zkvm_arch}, the left-side inputs: program code, commitments, and witness represent what the prover supplies to the zkVM, where commitments bind the prover to public artifacts, for example, committed program, committed state, or committed data roots, and the witness contains private execution details. The internal blocks represent the virtualized execution environment (memory, OS/runtime, registers, CPU/ISA interpreter), while the backend represents the proving backend that turns the execution trace into a proof. The right-side outputs are the proof and the program output that a verifier checks.
A key takeaway is that zkVM performance is usually dominated by trace generation and the proving backend kernels, not by the conceptual VM blocks themselves. In practice, the cost concentrates in constraint checks for the ISA semantics, memory consistency checks, hashing and commitment openings for trace integrity, and algebraic kernels such as MSM and FFT/NTT depending on the proof family.

\begin{figure}[!ht]
\centering
\resizebox{0.7\textwidth}{!}{%
\begin{circuitikz}
\tikzset{
  font=\Large,
  block/.style={draw, rounded corners=6pt, thick, minimum width=6.2cm, minimum height=0.95cm, align=center, fill=gray!8},
  io/.style={draw, rounded corners=6pt, thick, minimum width=3.6cm, minimum height=0.95cm, align=center, fill=gray!4},
  title/.style={font=\Huge},
  arrow/.style={->, >=Stealth, very thick}
}

% -- zkVM stack (center) --
\node[title] (t) at (0,6.6) {zkVM};

\node[block] (mem) at (0,5.3) {Main Memory};
\node[block] (os)  at (0,4.0) {Operating System};
\node[block] (reg) at (0,2.7) {Registers};
\node[block] (cpu) at (0,1.4) {CPU};
\node[block] (be)  at (0,0.1) {Backend};

% container outline
\draw[rounded corners=10pt, thick] (-3.5,5.95) rectangle (3.5,-0.55);

% -- Inputs (left) --
\node[io] (code) at (-6.4,5.3) {Program Code};
\node[io] (comm) at (-6.4,3.3) {Commitments};
\node[io] (wit)  at (-6.4,1.3) {Witness};

% -- Outputs (right) --
\node[io] (proof) at (6.4,4.4) {Proof};
\node[io] (out)   at (6.4,2.0) {Program Output};

% -- Arrows --
\draw[arrow] (code.east) -- (-3.5,5.3);
\draw[arrow] (comm.east) -- (-3.5,3.3);
\draw[arrow] (wit.east)  -- (-3.5,1.3);

\draw[arrow] (3.5,4.4) -- (proof.west);
\draw[arrow] (3.5,2.0) -- (out.west);

\end{circuitikz}
}
\caption{General zkVM Architecture}
\label{fig:zkvm_arch}
\end{figure}

The first step of most zkVMs is to receive an input program from a higher-level language such as C++, Circom \cite{circom}, Rust, or others. The program is then compiled to bytecode that the VM can translate to an instruction set architecture (ISA) that is generalized, like RISCV, or specialized, such as Miden Assembly \cite{polygonmiden}. These ISAs are minimal and optimized for cryptographic operations such as hashing to optimize the performance of the zkVM.
After the program is translated to bytecode, the setup phase involves tasks like concealing private inputs, initializing a polynomial commitment scheme, and key generation. Constraints are then applied to the program and witness to ensure computational integrity, aligning with the ISA. Each constraint corresponds to a polynomial for each statement in the execution trace. To check the validity of each instruction in the execution trace, a verifier can open any instruction in the zkVM to verify it was computed correctly.

The zkVM distinguishes itself from writing a specialized circuit for computations by its high optimization level and reduction of complexity from designing a circuit. In a ZK computation, three common overheads must be taken into account:

\begin{enumerate}
    \item  Range Checks: A range check is a way to prove that an element in a finite field falls within a specific interval \texttt{[a, b]}.
    \item Bitwise operations: Nearly every computation in a ZK context needs some bitwise operation for tasks such as hashing, and these operations need to be optimized for elements in finite fields.
    \item Hashing: Traditional hashes such as SHA256 often incur high overhead in many arithmetizations because they are built from bitwise operations (XOR/AND), rotations, and additions over words, which must be emulated using boolean constraints or non-native field arithmetic. In contrast, SNARK-friendly hashes such as Poseidon, Rescue, or MiMC are designed to align with finite-field operations, reducing constraint counts and improving prover time. Some zkVMs also isolate hashing into specialized components or choose trace layouts that make hash-heavy workloads cheaper to prove.
\end{enumerate}

\subsubsection{Projects}

Some of the prominent zkVMs over the last decade are TinyRAM \cite{tinyram}; Hawk \cite{hawk}; Zinc \cite{zinc}; RISC Zero \cite{risczero},  CairoVM \cite{starknetCairo}, Leo \cite{Aleo}, Miden \cite{polygonmiden}, Triton \cite{triton}; OlaVM \cite{olavm}; Powdr \cite{powdr}, Jolt \cite{jolt};  SP1 \cite{sp1}, Nexus \cite{nexus}, and Valida \cite{valida}.
Table~\ref{tab:zkvm_design_choices} summarizes representative design choices that recur in the zkVM ecosystem like arithmetization target, proof family, hash strategy, and recursion support.
Across many current systems, two recurring patterns appear: VM-style designs that prove an execution trace often favor transparent setups and hash-based commitments, frequently at the cost of larger proofs and hash-heavy traces; SNARK-style backends often emphasize smaller proofs and faster verification, but may rely on stronger setup assumptions depending on the concrete scheme. These tradeoffs matter directly for rollups, zkEVMs, and general verifiable compute.

\begin{table}[t]
\centering
\caption{Representative zkVM design choices.}
\label{tab:zkvm_design_choices}
\small
\renewcommand{\arraystretch}{1.1}
\setlength{\tabcolsep}{6pt}
\resizebox{\textwidth}{!}{%
\begin{tabular}{l l l l l}
\hline
\textbf{zkVM} &
\textbf{Execution encoding} &
\textbf{Proof family} &
\textbf{Hash strategy} &
\textbf{Recursion / composition} \\
\hline
RISC Zero &
RISC-V trace + constraints &
STARK (optional SNARK wrap) &
hash-heavy (SHA-2 common) &
supported \\
CairoVM &
AIR (trace constraints) &
STARK &
hash-heavy trace &
supported \\
Miden VM &
AIR (stack trace) &
STARK &
ZK-friendly perm hash &
supported \\
\hline
\end{tabular}}
\end{table}

A project in the space is RISC Zero's zkVM \cite{risczero} which allows users to demonstrate the accurate execution of arbitrary Rust and C++ code on an embedded RISC-V micro-processor. Starkware~\cite{starknetCairo}, the company behind zkSTARKs, developed CairoVM. CairoVM implements a STARK-based von-Neumann architecture, supporting traditional language features like conditional branching, function calls, recursion, and a restrictive memory model. Polygon's Miden VM~\cite{polygonmiden} is a zkVM designed for zero-knowledge rollups, which enhances Ethereum with features like parallel transaction execution and client-side proving.

\subsection{Domain Specific Languages}
\label{sec:DSLs}

\begin{table}[t]
\caption{Comparison of zkDSLs}
\centering
\resizebox{0.6\textwidth}{!}{%
\begin{tabular}{|>{\raggedright\arraybackslash}p{0.15\linewidth}|>{\raggedright\arraybackslash}p{0.15\linewidth}|>{\raggedright\arraybackslash}p{0.2\linewidth}|>{\raggedright\arraybackslash}p{0.15\linewidth}|>{\raggedright\arraybackslash}p{0.2\linewidth}|}
\hline
\textbf{zkDSL}            & \textbf{Proof System(s)}        & \textbf{Intermediate Representation}                         & \textbf{Syntax Comparison}   & \textbf{Purpose(s)}               \\ \hline
O1JS/Snarky & Kimchi                 & Plonk Variant& TypeScript          & General, Smart contracts \\ \hline
Circom      & Groth16                & R1CS                                                & N/A                 & Circuit development      \\ \hline
Cairo       & STARK                  & Algebraic Intermediate Representation& Rust                & Smart contracts          \\ \hline
Noir        & Plonk, Groth16, Marlin & Abstract Circuit Intermediate Representation& Rust                & Circuit development      \\ \hline
Juvix       & Plonk, Halo2           & VampIR                                              & Ocaml               & Anoma Intents            \\ \hline
Lurk        & Groth16, Nova          & R1CS                                                & Common Lisp, Scheme & General                  \\\hline
\end{tabular}
}
\label{tab:zkdsl_compare}
\end{table}

\subsubsection{Motivation and Definition}

zkDSL (zero-knowledge domain-specific language) bridges high-level program representation and the low-level details of ZKPs, like writing circuits. Its purpose is to convert high-level language into an arithmetic circuit, which can then be used in a proving system to produce a proof for a program's execution. Common circuit targets include R1CS, Plonk, and AIR. Each zkDSL has a specific focus; for instance, Circom~\cite{circom} specializes in writing arithmetic circuits, while Lurk~\cite{lurk} is a general-purpose programming language.
Table~\ref{tab:zkdsl_compare} gives a mapping from language frontends to the constraint targets and proof backends they commonly support.

\subsubsection{Methodology}

While zkDSLs offer abstraction, they have some drawbacks. They face memory management challenges due to the complexities of zero-knowledge proofs and cryptographic computations, requiring efficient handling of cryptographic primitives, large boolean circuits, and data structures. Additionally, zkDSLs struggle with conventional programming paradigms, such as recursion, traditional conditionals, mutable variables, and user-defined structures.
In practice, circuit and DSL ecosystems must defend against several common classes of bugs: (i) under-constrained circuits, where the constraint system does not fully enforce the intended relation and allows invalid witnesses; (ii) unsafe gadgets, where reusable components omit edge cases for example, missing constraints in arithmetic or comparator gadgets; (iii) range-check mistakes, where field elements are not properly bounded and can represent unintended values; and (iv) witness malleability patterns, where multiple witnesses satisfy the same public statement in unintended ways.
To mitigate these risks, projects increasingly use a mix of unit tests, property-based testing, fuzzing, differential testing between backends, and audit-driven checklists. Practitioner resources include public audit reports and issue trackers for major proving systems and toolchains, which provide concrete examples of real-world circuit failures and patch patterns.
Figure~\ref{fig:dsl_process} provides a compact view of how zkDSLs translate developer code into constraints and then into proofs. The main takeaway is that performance and security depend on both compilation choices (constraint generation) and backend choices (proving and commitment schemes).

\begin{figure}[t]
\centering
\resizebox{0.9\linewidth}{!}{%
\begin{tikzpicture}[
  node distance=2.3cm,
  >={Stealth},
  every node/.style={font=\large},
  proc/.style={
    draw, rounded corners=10pt, thick,
    minimum width=3.2cm, minimum height=1.05cm,
    align=center, fill=gray!8
  },
  arr/.style={->, very thick}
]
\node[proc] (code) {Code};
\node[proc, right=of code] (cs) {Constraint System};
\node[proc, right=of cs, minimum width=3.9cm] (cc) {Cryptographic\\Compiler};
\node[proc, right=of cc] (ps) {Proving System};
\node[proc, right=of ps] (proof) {Proof};

\draw[arr] (code) -- (cs);
\draw[arr] (cs) -- (cc);
\draw[arr] (cc) -- (ps);
\draw[arr] (ps) -- (proof);
\end{tikzpicture}%
}
\caption{Overview of zk-DSL Process}
\label{fig:dsl_process}
\end{figure}

\subsubsection{Projects}

To organize the zkDSL ecosystem more clearly, we group representative projects into three categories: (i) constraint-centric DSLs that prioritize explicit circuit construction (Circom), (ii) higher-level languages and VM-oriented stacks that compile programs into an execution trace or constraint IR (Cairo and Lurk), and (iii) compiler frameworks that target multiple constraint backends (CirC). We use these categories when relating tools to applications such as rollups, privacy protocols, and ZKML.
Circom \cite{circom} has been used in applications such as TornadoCash \cite{tcash} and DarkForest \cite{darkforest}. Circom is both a programming language and a compiler for the creation of arithmetic circuits, which are subsequently compiled into Rank-1 Constraint Systems (R1CS).
Circomlib \cite{circomlib}, an open-source library, provides reusable circuit templates. O1JS \cite{o1js} generates and validates zero-knowledge proofs from R1CS. Together, these tools abstract the low-level details of ZK proving, providing a higher-level interface for circuit construction.

Lurk \cite{lurk} is a Turing complete LISP-based programming language that autonomously generates zkSNARKS for arbitrary programs. It presents programs as data to the universal Lurk interpreter circuit to achieve Turing completeness without compromising the size of the proof artifacts generated.
Leo~\cite{leo} is a programming language for developing formally verified zero-knowledge applications on the blockchain. It offers an unrestricted execution environment with no limits on running time, stack size, or instruction sets, providing flexibility. Besides ensuring application privacy and reducing maximal-extractable value (MEV), Leo has two key features: formal verification of applications against high-level specifications and accessible succinct verification regardless of application size.

\subsection{Libraries and Frameworks}

\subsubsection{Motivation and Definition}
\begin{figure}[h!]
    \centering
    \includegraphics[width=0.7\linewidth]{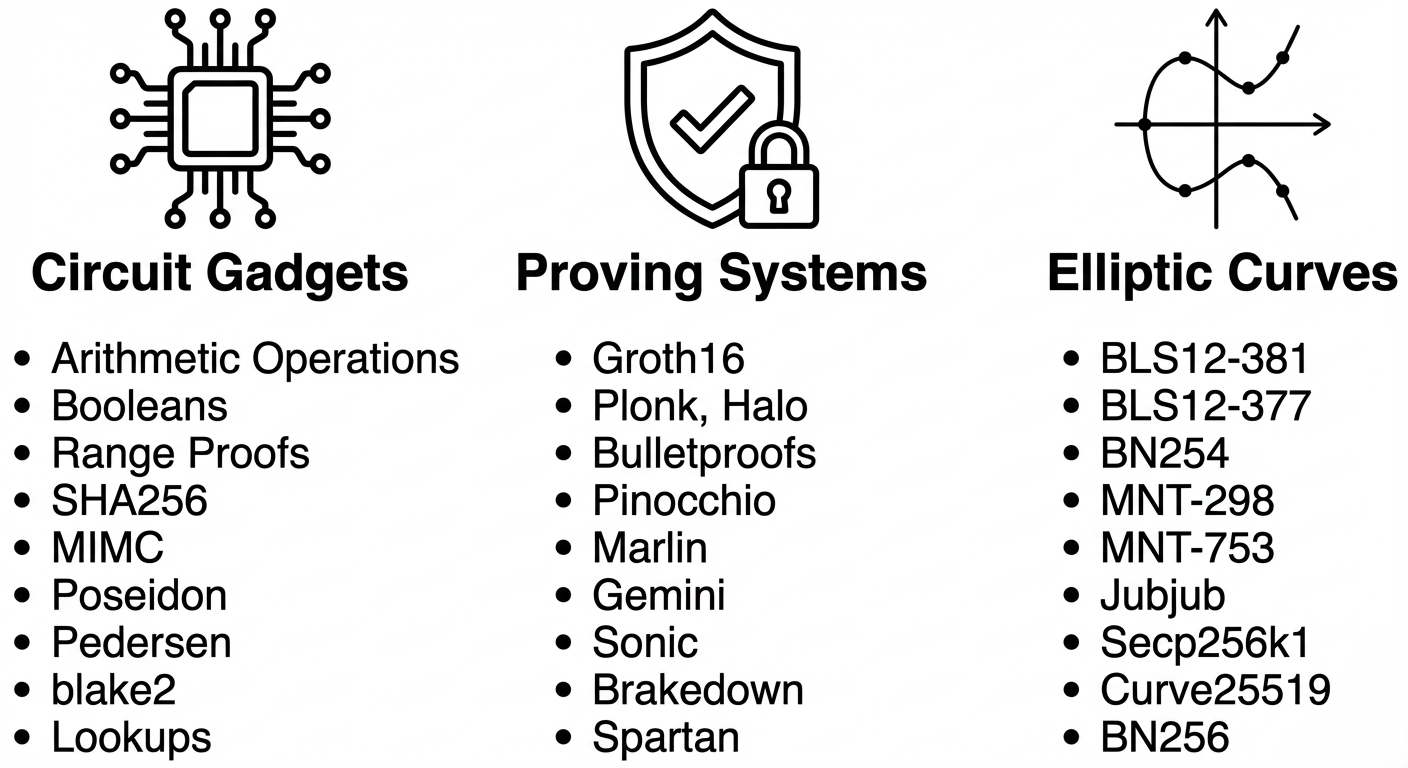}
    \caption{Categories of Cryptographic Components}
    \label{fig:cryptographic_components}
\end{figure}

In the domain of zero-knowledge proofs (ZKP), the development process has a high barrier to entry, requiring working knowledge of computational strategies, cryptographic primitives, elliptic curves, and ZKP protocol design. For developers, the hardship lies in crafting reusable code and mitigating redundancy, particularly within ZK proofs, where constraints such as circuit size and gas usage can impose significant limitations. To address these challenges, a multitude of libraries have emerged, offering implementations for cryptographic gadgets that facilitate the construction of Rank-1 Constraint Systems (R1CS) instances from modular gadget classes as listed in Figure~\ref{fig:cryptographic_components}. To facilitate the proving and verification process, elliptic curves are utilized. Operations on these curves is relatively standardized and does not always need to be recreated, so libraries avoid the issue of duplicated code in this area. The combination of circuit gadgets and elliptic curves allow for the modular composition of an application in conjunction with a compatible proving system such as Groth16 \cite{groth16} or Bulletproofs \cite{bulletproofs}.

These libraries abstract the complexities of preprocessing in a modular manner, letting developers focus on higher-level design and reuse code across ZKP applications.
A central concept used throughout libraries and backends is a \emph{commitment}. A commitment is a cryptographic binding to a value that can be opened later. Informally, it provides \emph{binding} (the committer cannot change the committed value after committing) and often \emph{hiding} (the commitment does not reveal the value). In ZK systems, commitments are used to bind the prover to large objects that the verifier cannot read in full, such as polynomials, Merkle roots, or witness encodings. This is why ``Commitments'' appears as an explicit input in Figure~\ref{fig:zkvm_arch}: the prover commits to artifacts that will later be checked via small openings rather than full disclosure.

\subsubsection{Methodology}

In the design of ZKP frameworks and libraries, several fundamental principles are leveraged to optimize performance and efficiency. One important design choice is the operation over small fields, as demonstrated by systems like ethSTARK \cite{ethSTARK} and Plonky2 \cite{plonky2}. Unlike traditional elliptic curve groups, which require large field sizes (e.g., 256 bits) for standard security levels, these systems utilize smaller prime fields (e.g., 64 bits). This approach capitalizes on the efficiency of small-field arithmetic, resulting in state-of-the-art proving performance. Additionally, using small-field elements reduces storage requirements, improving CPUs' cache efficiency. Another critical consideration is the flexibility in field selection. These schemes often use computationally structured fields, providing additional optimization opportunities. Finally, these frameworks and libraries prioritize using cheaper cryptographic primitives, ensuring cost-effective and efficient ZKP implementations \cite{zkbench}.

Abstracting away finite field operations necessitates the implementation of fundamental arithmetic operations, including addition, subtraction, multiplication, (modular) exponentiation, and inverse exponentiation, over finite fields in a modular and efficient manner. Efficient implementation of pairings, scalar multiplication, and multi-scalar multiplication (MSM) over pairing-friendly elliptic curves is also essential for achieving computational efficiency. Efficient implementation of these operations directly affects the performance of cryptographic protocols that rely on pairing-friendly elliptic curves. SNARKs can be inefficient with traditional hash algorithms and signature schemes, like SHA-2 and ECDSA, for example due to non-native field arithmetic. To address this, SNARK-friendly alternatives such as Poseidon Hash, Keccak Hash, Pedersen Hash, MIMC Hash, and Ed25519 (EdDSA signature) have been proposed, optimized for SNARKs \cite{zkbenchpaper}.

Concrete optimization themes that repeatedly yield large speedups in practice include: (i) MSM improvements (for example, Pippenger-style bucket methods, endomorphism-based speedups, and better batching), (ii) FFT/NTT tuning (for example, domain selection, parallelization, and cache-aware layouts), (iii) lookup arguments and custom gates that reduce constraint counts for common primitives, (iv) recursion and proof composition that amortize verification across many steps (for example, rollups and IVC), and (v) hardware acceleration that targets MSM/NTT/hashing kernels.

\begin{table}[!ht]
\caption{Popular Elliptic Curves in ZK Libraries}
\centering
\resizebox{0.4\textwidth}{!}{%
\begin{tabular}{|c|c|c|c|}
\hline
              & BN254                     & BLS12\_381                & BLS12\_377                \\ \hline
arkworks      & \checkmark& \checkmark& \checkmark\\ \hline
gnark         & \checkmark& \checkmark& \checkmark\\ \hline
blstrs        &                           & \checkmark&                           \\ \hline
ffjavascript  & \checkmark& \checkmark&                           \\ \hline
pairing\_ce   & \checkmark& \checkmark&                           \\ \hline
zkcrypto      &                           & \checkmark&                           \\ \hline
halo2\_curves & \checkmark&                           &                           \\ \hline
pasta\_curves &                           & \checkmark&                           \\ \hline
\end{tabular}
}
\label{tab:my_table}
\end{table}

Table~\ref{tab:my_table} focuses on elliptic-curve support that is central to many SNARK libraries. In contrast, STARK-style systems often rely on hash-based polynomial commitments and Merkle-based authentication rather than pairing-friendly curves, which is one reason their setup models and performance profiles differ from pairing-based SNARK families.
A recursive SNARK is a proof system where the statement being proved includes the correctness of verifying another proof. Concretely, recursion proves that a verifier circuit accepts a previous proof together with a new computation step, allowing many steps to be composed into a single succinct proof. This enables incremental verifiable computation (IVC) and constant-size proofs for long-running computations, and it is widely used in rollups, proof-carrying data, and systems that require repeated updates with bounded verification cost.

\subsubsection{Projects}

Arkworks \cite{arkworks} is a Rust ecosystem tailored for zkSNARK programming, offering a set of libraries to streamline zkSNARK application development. These libraries implement essential components, including generic finite fields and R1CS constraints, enabling developers to seamlessly integrate zkSNARK functionality into their applications.
Gnark \cite{gnark-v0.9.0} is a library that enables developers to design circuits in Go. It employs a versatile API and command line interface to accomplish the ZKP process in a familiar way to developers. The two proving schemes that Gnark supports are Groth16 and Plonk, along with a variety of elliptic curves for developers to choose from, such as BN254, BLS12-381, and BLS12-377. A core focus of the Gnark library is speed of the prover and verifier, since it offers an API for both the frontend and the backend of the proving process.
CirC \cite{circ} is a project dedicated to compiler infrastructure for cryptosystems and verification. It focuses on cryptographic tools such as proof systems, multi-party computation, and fully homomorphic encryption, typically applied to computations expressed as systems of arithmetic constraints. These applications require compilers that can translate high-level programming languages (e.g., C) into such constraints. CirC aims to provide a shared infrastructure for building constraint compilers, offering a valuable resource across various applications. These compilers can translate code into a Rank-1 Constraint System (R1CS), enabling efficient implementation of cryptographic protocols and verification mechanisms.

The TypeScript library O1JS \cite{o1js} is designed to cater to users with a web development background, offering a user-friendly approach to writing ZK programs and smart contracts for the Mina \cite{Mina} blockchain. It is described as an embedded DSL and is executed as normal TypeScript. This library permits developers to write arbitrary ZK programs utilizing many built-in provable operations, including basic arithmetic, hashing, signatures, boolean operations, and comparators.
Various implementations of Zcash's Halo2 \cite{halo2zcash} proving system, such as those by Axiom and the Ethereum Foundation, provide fundamental primitives for writing zero-knowledge proof circuits. The proving system involves several stages, from committing to polynomials that encode the main components of the circuit, including cell assignments, permuted values, products for lookup arguments, and equality constraint permutations.
Nova \cite{nova} is a recursive SNARK that enables incrementally verifiable computation (IVC), a cryptographic primitive that allows a prover to produce a proof of the correct execution of a long-running sequential computation incrementally. IVC allows proofs to build on each other efficiently, speeding up the entire proving and verification process.

\subsection{Hardware Acceleration}

\subsubsection{Motivation and Definition}

Historically, the speed and memory requirements of ZK proof generation have limited their applicability. The required computations inside of a ZKP, such as hashing, multi-scalar multiplications, and fast-fourier transforms, create a burden for each use case. To reduce the overhead required by ZKPs, various projects have emerged to enhance the performance of ZKPs and their potential implementations.

\subsubsection{Methodology}

Hardware acceleration is the use of optimized or dedicated computer components to improve the performance and efficiency of a specific operation. The main instruments used to accelerate this in the ZK world are field-programmable gate arrays (FPGAs), graphics processing units (GPUs), and application-specific integrated circuits (ASICs). The limiting factors for hardware acceleration projects are memory capacity, memory access speed, data transfer speed, and arithmetic unit speed. In proof systems where both number theoretic transforms (NTTs) and multi-scalar multiplications (MSMs) are used, the majority of the proof generation time is spent on MSMs, with NTTs accounting for the remainder. Both MSMs and NTTs present performance challenges that can be addressed in several ways. MSMs can be executed across multiple threads, enabling parallel processing. However, when performing large vector operations, multiplications may still be slow and require considerable memory. Additionally, MSMs face scalability issues and can remain sluggish even when extensively parallelized. On the other hand, NTTs involve frequent data shuffling during the algorithm, making them difficult to distribute across a computing cluster. They also require significant bandwidth when run on hardware due to the need to load and unload elements from large datasets.

Both MSM and NTT can be accelerated on GPUs, particularly MSM using the Pippenger algorithm. This process involves rewriting the computationally intensive tasks from the CPU to the GPU using CUDA or OpenCL, allowing the code to be compiled and executed directly on the GPU. For finer-grained acceleration, developers can optimize memory usage by maximizing fast memory and minimizing slow-access memory to reduce costly data transfers, especially between the CPU and GPU. Additionally, optimizing execution configuration by balancing work across multiprocessors, building concurrent kernels, and allocating resources judiciously can maximize hardware utilization. The goal is to parallelize the entire workflow, minimizing sequential execution where different parts depend on each other's results. Open-source implementations allow developers to quickly start their modifications \cite{paradigmzkhardware}.
FPGAs, or field-programmable gate arrays, offer programmable hardware fabric that can be reconfigured multiple times, cutting manufacturing costs compared to ASICs and providing greater flexibility in hardware resource usage than GPUs. Although optimizing NTTs on GPUs is achievable, frequent data shuffling can lead to significant communication overhead between the GPU and CPU. By implementing the logic directly into the circuit design, FPGAs can potentially perform the task faster. Most open-source implementations for zero-knowledge proofs are written in Rust due to its memory safety and cross-platform compatibility. However, FPGA development tools are typically adapted to C/C++, requiring teams to translate these implementations \cite{revisitingzkhardware}. Table~\ref{tab:hw_compare} provides a compact CPU/GPU/FPGA/ASIC comparison across the dominant proving kernels.
Acceleration changes which applications are feasible. High-throughput rollups, zkEVM-style proving, and large ZKML inference proofs often become practical only when MSM/NTT/hashing kernels are accelerated. However, even with acceleration, systems can remain bottlenecked by memory footprint, proof size, and data-availability costs in blockchain deployments.

\begin{table}[t]
\centering
\caption{Hardware acceleration comparison for common ZK kernels.}
\label{tab:hw_compare}
\resizebox{0.95\textwidth}{!}{%
\begin{tabular}{l c c c c}
\hline
\textbf{Kernel / constraint} & \textbf{CPU} & \textbf{GPU} & \textbf{FPGA} & \textbf{ASIC} \\
\hline
MSM & baseline, flexible & strong parallelism & low-latency pipelines & highest throughput potential \\
FFT/NTT & optimized libraries & good if bandwidth fits & strong if streaming fits & strong if specialized \\
Hashing / Merkle & baseline & strong for batched hashing & very strong for pipelines & very strong for fixed hash \\
Memory bandwidth & cache-limited & HBM helps, PCIe transfers hurt & on-chip bandwidth helps & design-specific \\
\hline
\end{tabular}}
\end{table}

\subsubsection{Projects}

Ingonyama \cite{ingonyama} is a hardware acceleration company that integrates chip design with mathematics and advanced algorithms to enhance the performance of compute-intensive cryptography. They maintain a library called ICICLE, a cryptography library for ZKPs using GPUs. ICICLE implements various cryptographic primitives such as elliptic curve (EC) operations, multi-scalar multiplication (MSM), number theoretic transform (NTT), and the Poseidon hash on GPUs.
Cysic \cite{cysic} is a ZK accelerator focused on developing ASICs and their accelerated zkVM. The system architecture features an executor responsible for executing programs, hardware for controlling and distributing segments, and a configurable number of specialized chips to generate ZK proofs for each segment program.
Fabric Cryptography \cite{fabric} introduced The Fabric Verifiable Processing Unit (VPU), a processor designed for cryptography applications ranging from ZKP to FHE. The VPU features a custom instruction set architecture tailored for next-generation cryptography, including ZKP, FHE, MPC, and other algorithms. It offers acceleration for MSM, NTT, polynomial evaluation, as well as Poseidon, Blake, and other hash functions. The VPU supports multiprecision vector lanes up to 384-bit and includes a RISC-V core for enhanced programmability.

Irreducible \cite{Irreducible} offers proof-as-a-service, designed for scalability and powered by FPGA-accelerated server clusters. Irreducible supports popular proof systems such as Plonky2 and Polygon zkEVM, with plans to support next-generation systems like Binius and Plonky3. Their FPGA-accelerated server clusters are specifically designed for cryptographic computation at scale.
Supranational \cite{Supranational} offers hardware-accelerated cryptography for verifiable and confidential computing. They provide BLST, an IETF-compliant BLS12-381 signature library focused on security and performance.

\section{Blockchain Applications} \label{sec:applications}
\begin{table}[!ht]
\caption{Overview of blockchain-based applications of Zero-Knowledge Proofs}
\centering
\resizebox{0.6\textwidth}{!}{%
\label{tab:use_case_projects}
\begin{tabularx}{\textwidth}{|l|X|}
\hline
\textbf{Use Case} & \textbf{Projects} \\
\hline
Layer 1 Blockchains & ZCash, Aleo, Mina \\
\hline
Layer 2 Scaling & Polygon zkEVM, zkSync Era, Scroll, Linea, Starknet, Aztec \\
\hline
Smart Contract/Transaction Privacy & Hawk, Tornado Cash, Privacy Pools, Penumbra, Mina zkApps, Noir \\
\hline
Proof of Identity & Semaphore, WorldID, zPass, Galxe protocol \\
\hline
Supply Chain/Enterprise Blockchain & QEDIT, zk-BeSC, Sahai et al. \\
\hline
Interoperability & zkBridge, Telepathy, \\
\hline
Blockchain Storage & Herodotus, FileCoin \\
\hline
Proof of Reserves & Provisions: Privacy-preserving proofs of solvency for Bitcoin exchanges, Proven: ZeKnow Solv \\
\hline
\end{tabularx}}
\end{table}

\subsection{Layer 1 Blockchains}

\begin{figure}[!ht]
    \centering
    \includegraphics[width=0.75\linewidth]{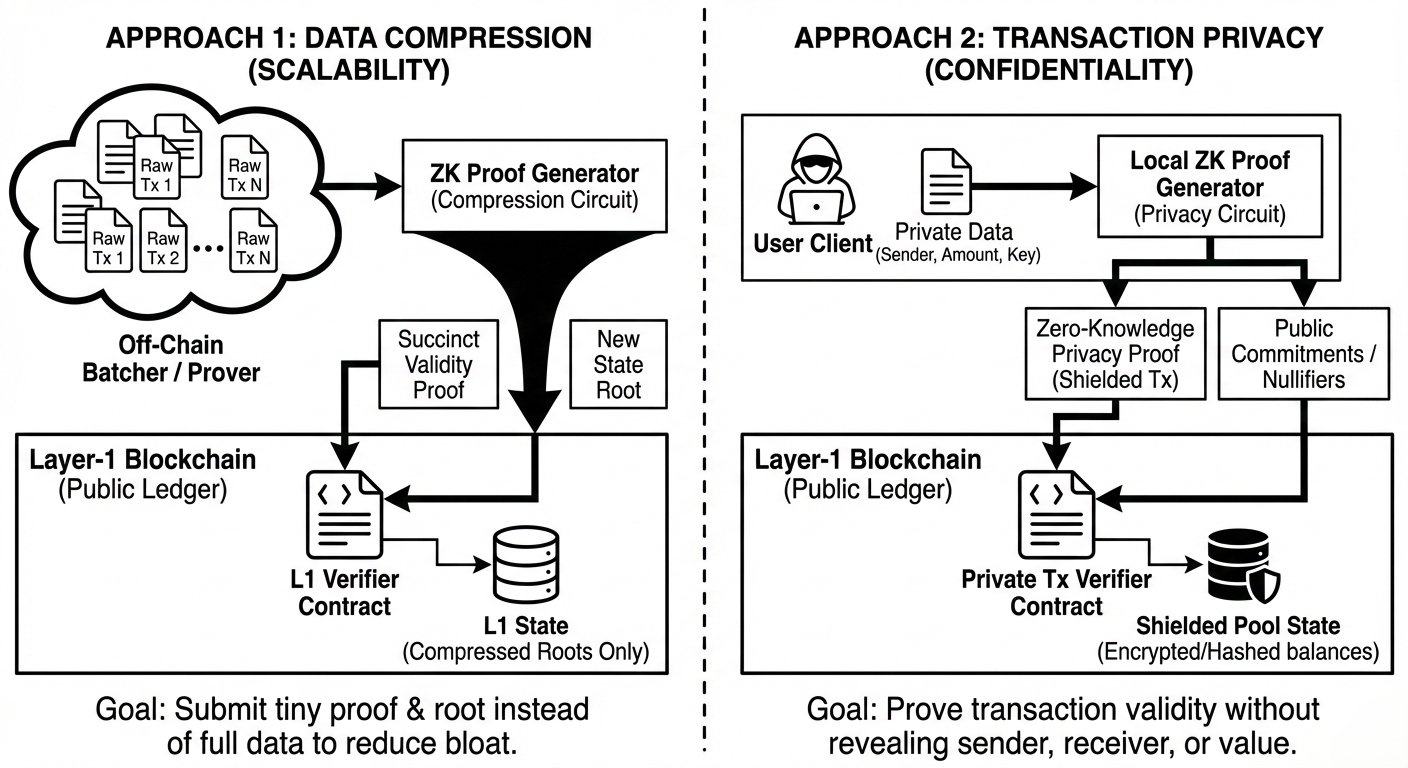}
    \caption{Comparison of Layer-1 Data Compression and Transaction Privacy Approaches using ZKPs}
    \label{fig:layer1}
\end{figure}

\subsubsection{Motivation and Definition}

Layer 1 blockchains, such as Ethereum and Bitcoin, are foundational blockchain networks that provide a transparent and immutable ledger for data storage. However, these platforms face challenges in privacy and scalability. The emergence of zero-knowledge proofs (ZKPs) has induced the development of novel Layer 1 solutions that prioritize privacy and data efficiency, as depicted in Figure~\ref{fig:layer1}. These blockchains integrate ZKPs directly into their base layers, offering a dual advantage: enhanced privacy through transaction concealment and improved scalability via data compression. In practice, privacy-first Layer 1s select a proving system based not only on proof size and prover cost, but also on the setup model (trusted vs.\ transparent) and whether recursive composition is a core requirement for keeping verification lightweight over long histories. The primary motivation for ZKP-based Layer 1 blockchains is to reconcile the inherent transparency of conventional blockchains with the increasing demand for data privacy and efficient on-chain data management.

\subsubsection{Methodology}
Layer-1 blockchains utilize ZKPs to obscure critical transaction details, such as the identities of parties involved and the transaction amounts. This concealment is achieved through advanced cryptographic techniques, including zkSNARKs or zkSTARKs, that validate transactions without revealing their underlying data. To address scalability challenges, ZKP-based Layer-1 blockchains employ state compression. This technique uses ZKPs to produce compact proofs that verify large sets of transactions or state transitions, thereby reducing the on-chain data volume required. In these networks, participating nodes are responsible for generating and verifying ZKPs. This ensures transaction integrity while maintaining privacy. The consensus mechanisms of these blockchains are uniquely designed to incorporate ZKP validation. This integration ensures that only transactions authenticated through ZKPs are confirmed and appended to the blockchain. However, these benefits introduce operational complexity, such as user UX can degrade due to wallet and key-handling requirements, proving latency can impact transaction confirmation time, especially on constrained devices, and key management becomes a first-class security concern, e.g., circuit upgrades, verification keys, and setup artifacts when applicable.

\subsubsection{Applications}

ZCash \cite{ZCash}, one of the earliest privacy-focused Layer 1 blockchains, uses zkSNARKS to enable private transactions, where the identities of the sender and receiver, and the transaction amounts, are encrypted. In ZCash, zkSNARKS enable the encryption of transaction data, ensuring the anonymity of both the sender and receiver, as well as the confidentiality of the transaction amount. This mechanism allows the network to validate transactions without revealing the underlying data, thereby preserving users' privacy. ZCash's implementation of zkSNARKS is notable for its efficiency, allowing for the verification of transactions in a matter of milliseconds, which is a significant advancement over other cryptographic techniques that were more computationally intensive. ZCash uses shielded transactions, in which transaction metadata is encrypted and zkSNARKS are used to prove compliance with consensus rules without revealing the underlying data. This approach enables a high degree of privacy while maintaining the integrity and security of the blockchain. ZCash offers optional privacy features allowing users to choose between shielded and transparent transactions, unlike Aleo, where privacy is mandatory for all transactions.

Aleo \cite{Aleo} is a distinctive Layer 1 blockchain platform that employs zero-knowledge proofs to enhance privacy and scalability in decentralized applications (dApps). It achieves this through a zkSNARK framework for constructing dApps that can perform computations in a private and verifiable manner.
The Mina Protocol \cite{Mina} is a Layer 1 blockchain that addresses scalability and privacy through recursive zkSNARKS. This protocol maintains a very small constant-size proof of chain validity through recursive composition, regardless of the total number of transactions processed. This is achieved through the recursive composition of zkSNARKS, which allows each new block to contain proof of the validity of the entire blockchain history. As a result, the Mina Protocol maintains a drastically reduced blockchain size compared to traditional blockchains, enhancing scalability and usability.

\subsection{Layer 2 Scaling}

\subsubsection{Motivation and Definition}
As mentioned above, ZKP can be used for data succinctness, especially for blockchain validity proofs. A significant pain point for blockchains is the scarcity and cost of block space, which can be addressed through Layer 2 scaling solutions such as rollups. The most prominent types of rollups are optimistic rollups and ZK rollups, but we will focus on the latter. More precisely, validity rollups use succinct proofs to attest to correct state transitions: zero-knowledge privacy is optional and protocol-specific. In ZK rollups, zero-knowledge proofs are used to succinctly prove the validity of state changes to a Layer 1 blockchain without requiring validator nodes on a Layer 1 chain to execute those transactions. In essence, the Layer 2 rollup becomes a cheap, modular execution layer that benefits from the security and decentralization of Layer 1, enabling blockchains to scale by significantly reducing operational costs.

\subsubsection{Methodology}

In Layer 2 rollups \cite{efZKR}, all of which achieve finality on Ethereum Layer 1, heavy computation is handled outside the chain, and only state changes, deposits, and withdrawals are posted to Layer 1 through the rollup smart contract. This Layer 1 smart contract maintains a state root, the Merkle tree root of the rollup's state, including accounts and balances. A rollup sequencer is an entity that could range from a single server to a decentralized network of nodes that orders transactions,  produces L2 blocks, and adds rollup transactions to the ZK-rollup contract with a ZK validity proof. The sequencer publishes a batch of highly compressed transactions, along with the previous state root and the new state root, after processing all transactions in the batch. Using the ZK validity proof provided by the sequencer, the rollup contract checks that the new state root is valid, then swaps the old root for the new one. The transaction batches published by sequencers are written to Ethereum in the form of encoded function calls, stored either in the calldata of the EVM, which is a data area used to pass arguments to a function and does not modify the blockchain's state, or in temporary data storage locations called blobs (post-EIP-4844). This provides a cheap way to store data on-chain, enabling individuals to reconstruct the rollup's state from these compressed transactions. Since the Layer 1 rollup contract can quickly verify a zk-SNARK or zk-STARK proof on any amount of large computation, computation is almost fully offloaded to the Layer 2. Through these methodologies, ZK rollups minimize the space and computation restraints of Layer 1 Ethereum, which simply validates state changes and provides inherent decentralization and security \cite{efZKR}. In practice, user fees are often dominated by (i) on-chain verification cost (proof verification and any aggregation), and (ii) data availability cost, where calldata and blob-based DA trade off pricing and throughput. Consequently, DA policy (rollup vs.\ validium vs.\ volition) and compression strategy can matter as much as the prover itself.

Since Layer 2 scaling solutions are built atop smart contract-enabled blockchains, most notably Ethereum rollups, they must submit proofs of their state transitions to the Ethereum Virtual Machine (EVM), which serves as Ethereum's Turing-complete computation engine. Since ZK-rollups must attest to the correctness of computations using ZKPs before posting batches to Layer 1, they must take additional steps during code execution, whereas the EVM does not. Rather than loading smart contract bytecode into the execution layer and simply posting the result of those computations, ZK-rollups must generate validity proofs for each transaction's state transition.
Because EVM opcodes are designed for general-purpose computations, proving EVM computations in ZK circuits is too resource-intensive and complex. Consequently, it is very difficult to ensure EVM-compatibility in ZK rollups, resulting in varying architectural designs:

\begin{itemize}
    \item zkEVM: A solution that embeds zero-knowledge proofs into EVM smart contract execution by recreating existing EVM opcodes for proving in circuits. It computes state transitions just like the typical EVM. However, it creates ZK validity proofs to verify the correctness at every operation, including state changes and computations.
    \item Custom VMs: Some approaches involve creating new high-level or intermediate languages and virtual machines that are more amenable to ZK proofs while still supporting EVM-like operations.
\end{itemize}

zkEVMs enable easier execution of Solidity smart contracts, whereas custom VMs may require developers to write smart contracts in another smart contract language or modify their existing EVM-based implementations due to ZK limitations.

\begin{table}[t]
\centering
\caption{Comparison of representative Layer 2 validity systems.}
\label{tab:l2_comparison}
\small
\renewcommand{\arraystretch}{1.1}
\setlength{\tabcolsep}{5pt}
\resizebox{\textwidth}{!}{%
\begin{tabular}{l l l l l l l}
\hline
\textbf{L2} & \textbf{Proof family} & \textbf{Setup} & \textbf{EVM level} & \textbf{Language/tooling} & \textbf{DA mode} & \textbf{Indicative proving throughput} \\
\hline
zkSync Era & SNARK & trusted & EVM-compatible & Solidity/Vyper + custom compiler & rollup/validium & medium \\
Polygon zkEVM & SNARK & trusted & EVM-equivalent & Solidity (no custom compiler) & rollup/validium & medium \\
Scroll & SNARK & trusted & EVM-compatible & Solidity (minimal changes) & rollup & medium \\
Linea & SNARK & trusted & EVM-equivalent & Solidity (standard tooling) & rollup & medium \\
StarkNet & STARK & transparent & custom VM & Cairo toolchain & rollup/validium/volition & high \\
Aztec & SNARK & trusted & custom VM & Noir/ACIR toolchain & rollup & low--medium \\
\hline
\end{tabular}}
\end{table}

\subsubsection{Applications}
Table~\ref{tab:l2_comparison} summarizes representative design points to aid interpretation of the following systems:
zkSync Era \cite{zkSync} is a zkEVM rollup developed by Matter Labs. Identical to the previously mentioned methodology, zkSync Era utilizes zkSNARKs to provide validity proofs of off-chain computation. Smart contracts can be written in Solidity or Vyper and called using the same clients as other EVM-compatible chains, thus making zkSync EVM-compatible. The zksolc compiler used on Era generates bytecode with optimizations in order to make operations more amenable to proof generation, using LLVM as an intermediate representation before executing zkEVM assembly code. Consequently, there are multiple differences from Ethereum, with many EVM opcodes having modified implementation rules. zkSync Era is an early EVM-compatible chain to implement native account abstraction, a system of smart-contract-based accounts with arbitrary logic first introduced in EIP-4337. Thus, every user account on zkSync Era can use smart accounts with their existing externally owned accounts (EOAs).

Matter labs also provides an open framework for deploying additional modular chains similar to zkSync Era called the ZK stack \cite{zkSync}. A modular chain from the ZK stack, called a hyperchain, runs a separate instance of the zkSync zkEVM and settles transactions on Ethereum's Layer 1. Hyperchains are linked via Hyperbridges, enabling asset transfers. While anyone can deploy Hyperchains, ensuring trust and full interoperability requires using the zkEVM engine from the ZK Stack, which powers zkSync Era. This uniformity in ZKP circuits across Hyperchains guarantees inherited security from Layer 1 without additional trust assumptions. Hyperchains implement a modular approach, allowing developers to choose or create their own blockchain components, except for the zkEVM core. Various options for sequencing transactions are available, ranging from centralized to decentralized sequencers to external protocols for customizing Hyperchain sequencing. Each Hyperchain can also manage its data availability (DA) policy, for example, a ``Validium'' architecture which stores state data off-chain rather than posting the calldata to L1, providing flexibility tailored to specific needs.

Polygon zkEVM \cite{polygon}, built by Polygon Labs, aims to offer full EVM-equivalency, with no separate compiler. Thus, ZK proving circuits verify most EVM opcodes as they are, with a few that carry minor differences that do not affect the developer experience. Because of this inherent equivalence and support for EVM opcodes, developers can deploy their existing L1 smart contracts directly to the Polygon zkEVM rollup with no necessary tweaks. Much like the zkSync stack, Polygon has the Chain Development Kit (CDK) \cite{polygon2}, which allows developers to deploy application-specific chains as validiums using the Polygon zkEVM.
Scroll \cite{scroll} represents another approach within the zkEVM landscape, focusing on EVM compatibility with necessary adaptations for zero-knowledge proofs. Scroll's zkEVM modifies certain EVM opcodes to fit the ZK-proof framework while maintaining the ability for developers to write and deploy Solidity contracts with no custom compiler. Although these modifications alter how some operations are handled compared to Ethereum, they are clearly documented, allowing developers to account for them during smart contract development. The tailored modifications aim to preserve the core experience of Ethereum smart contract interaction within the constraints of a ZK rollup environment.

Linea's zkEVM \cite{linea}, developed by ConsenSys for the Linea L2, closely mirrors the Polygon zkEVM, providing an EVM-equivalent experience without requiring a custom compiler. Supporting Solidity compilers, Linea enables developers to use well-known Ethereum tools like Hardhat and Foundry seamlessly. This compatibility eases the developer transition to Linea, with minor considerations for Solidity version compatibility. Additionally, Linea integrates the Canonical Message Service, a system allowing arbitrary data transfer between Linea and other networks, enhancing cross-chain communication and utility.

StarkNet \cite{starknetCairo}, built by StarkWare, differs from other ZK-rollups in that it uses STARK proofs rather than SNARK-based schemes. StarkNet uses Cairo, a specialized programming language, rather than Solidity. Cairo programs are compiled into Sierra, a safe intermediate representation, and subsequently into Cairo assembly (Casm) for execution by the StarkNet OS virtual machine. This two-step compilation process translates smart contract execution into the polynomial constraints required by STARK proofs, which validate StarkNet's block execution. Like zkSync, StarkNet adopts a native account-abstraction model in which accounts are smart-contract-based rather than traditional EOAs \cite{starknetAA}.

StarkEx \cite{starkEx}, also developed by StarkWare, is a specialized Layer 2 scalability service utilizing STARK proofs for high-throughput, low-latency applications on Ethereum. Unlike StarkNet, StarkEx is not a standalone blockchain but a service tailored for specific use cases, such as decentralized exchanges (DEXs) and NFT platforms. It allows these applications to define their own logic off-chain and post transactions to the service, which then generates STARK proofs attesting to the validity of transaction batches. These proofs are submitted and verified on L1. StarkEx also offers various data availability modes: ZK-Rollup, Validium, and Volition. This flexibility allows applications to optimize with any mix of on-chain and off-chain components.

Aztec Network \cite{aztecDocs} is a Layer 2 rollup which focuses on privacy preservation through the Noir programming language~\cite{noirLang}. Smart contracts in Aztec using Noir can include both public and private elements. These contracts are defined as sets of functions, both public and private, written as Noir circuits. Each function, representing a zk-SNARK verification key, operates on the contract's public and private state. In Aztec, the sequencer aggregates transactions into a block, generates state update proofs, and posts them to the Ethereum rollup contract. This architecture, while similar to other Layer 2 networks, differs notably in how it handles private state. The private execution environment within Aztec safeguards sensitive operations and data, ensuring that private information remains confidential \cite{aztecDocs}.

\subsection{Blockchain Interoperability}

\subsubsection{Motivation and Definition} With a highly fragmented landscape of different blockchain technologies, including Layer 1 chains and additional modular layers built atop them, there has arisen the need for composability among blockchains. For example, a significant pain point in the blockchain user experience is the difficulty of bridging tokenized assets from one blockchain to another. This problem extends past cross-chain transactions and financial asset liquidity, as it also concerns general message passing and data storage across fragmented blockchain networks. Even for Layer 2 rollups built atop the same Layer 1, each rollup's state constitutes a separate data moat, resulting in the same fragmentation problem even with shared transaction finality and consensus security. Cross-chain composability, also known as blockchain interoperability, has produced several approaches, including those that leverage ZKPs for succinct verifiable computation.

\subsubsection{Methodology} Zero-knowledge proofs are used to verify the occurrence of a state change or block execution on one chain to another chain. This property enables protocol developers to coordinate application logic across multiple blockchains, allowing users to instantly transact and pass data between networks. Typically, middleware generates validity proofs that are verified on the receiving blockchain, where a smart contract verifies the proof and executes the corresponding application logic according to the application's specification. For example, asset bridges lock up tokens on one chain and mint them on another, allowing users to transfer liquidity between chains. Further use cases include cross-chain DAO voting and Non-Fungible Tokens (NFTs).

\subsubsection{Applications}

\begin{figure}[!ht]
    \centering
    \includegraphics[width=0.75\linewidth]{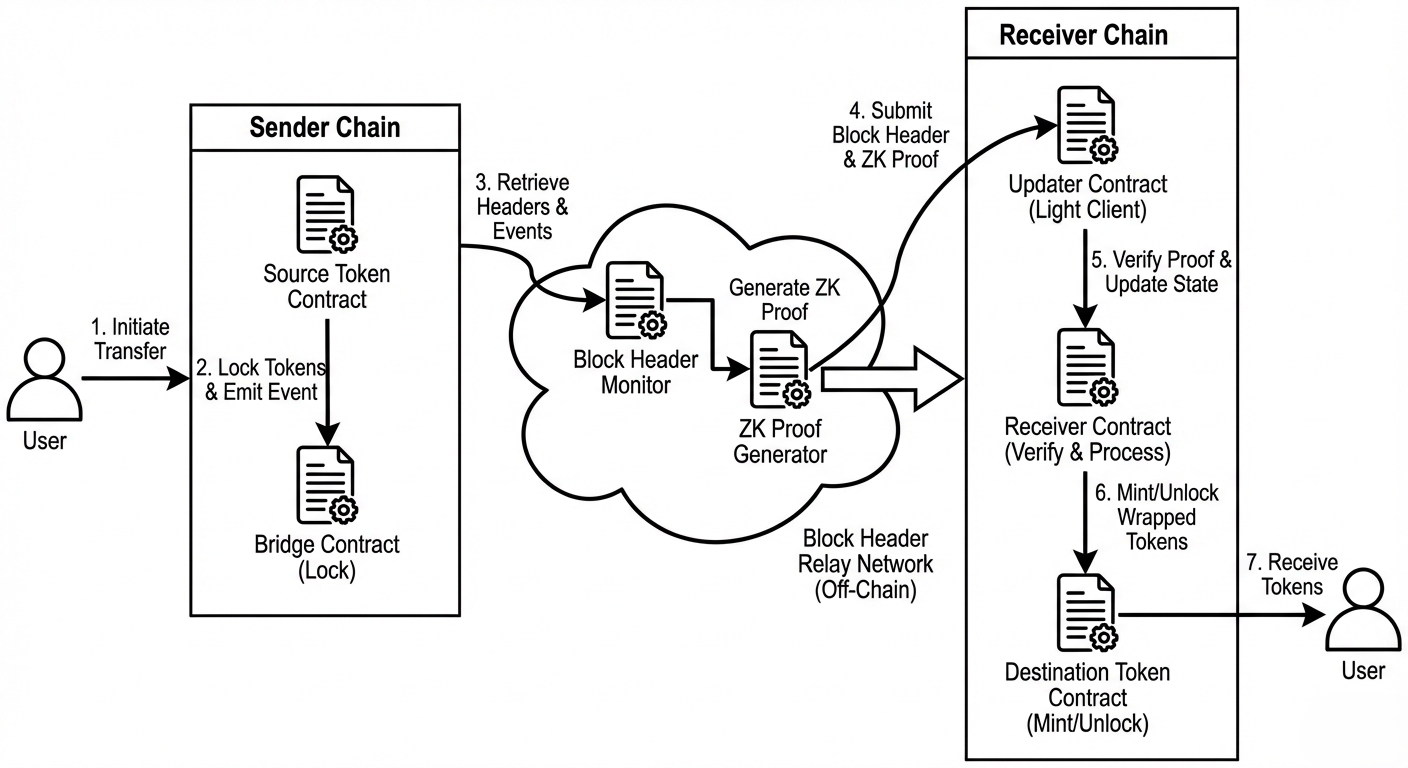}
    \caption{zkBridge architecture for a cross-chain token transfer}
    \label{fig:zkBridge}
\end{figure}

The zkBridge protocol \cite{xie2022zkbridge} illustrated in Figure~\ref{fig:zkBridge}, operates through a modular design that separates application-specific logic from the core functionality of relaying block headers. This core functionality is provided by a block header relay network, which is trusted only for liveness. This network relays block headers from one blockchain, along with zkSNARK correctness proofs, to an updater contract on another blockchain. The updater contract is responsible for maintaining a list of recent block headers from the sender chain, verifying proofs submitted by relay nodes, and updating the list accordingly. On the receiver blockchain, the updater contract provides an application-agnostic API that enables application smart contracts to obtain the latest block headers of the sender blockchain. This enables them to build application-specific logic on top of this information. Applications utilizing zkBridge generally deploy a pair of contracts: a sender contract on blockchain 1 and a receiver contract on blockchain 2. The receiver contract can call the updater contract to access block headers of blockchain 1, which it can then use to execute application-specific tasks.

Telepathy\cite{telepathy} is an interoperability protocol that allows arbitrary message passing between Ethereum and other chains. For developers that want to send a cross-chain transaction, they call the Telepathy router contract on Ethereum and must wait until the transaction reaches finality. To verify that a block has been finalized, an off-chain component called the Telepathy operator utilizes a zk-SNARK that proves the block header has signatures from a large percentage Ethereum validators. This proof is passed to the Telepathy light client contract on the destination chain, which verifies proofs and provides access to Ethereum's block headers. The Telepathy relayer accesses that light client data and generates a Merkle proof on the block to verify that the transaction exists and reached finality on Ethereum. This Merkle proof is passed to the Telepathy receiver contract, which verifies it and relays the smart contract call to the receiving contract on the destination chain, which executes the corresponding application logic as specified by the developer.

\subsection{Blockchain Storage}
\subsubsection{Motivation and Definition}Blockchain storage is essentially a way to save data in a decentralized network using the properties of the blockchain. In order to protect the saved data, blockchain uses data structures such as Merkle trees and Merkle Patricia trees. What is special about these data structures is that specific proofs can be constructed to show that a particular piece of data is contained in the structure and that, once a single piece of data is changed, the entire structure changes drastically. This essential property of blockchain ensures data integrity and invariance. However, as the quantity of data contained in a proof rises, so does the size of the proof. As a result, the cost of validating such proofs on the chain rises, rendering ordinary inclusion proofs economically unsustainable in most circumstances. A more scalable option is now preferred, in which the blockchain does not have to store large amounts of data but instead only small references to data stored on off-chain platforms. Zero-knowledge proofs make this possible: data and computations can be stored off-chain, and ZK Proofs can be used to succinctly and efficiently communicate a summary of these operations to the main chain without trust.

\subsubsection{Methodology}Zero-knowledge proofs could be used to minimize the cost of activities associated with verifying the inclusion of data in massive datasets and to validate that the procedure was completed correctly. The prover carries out the necessary calculation and generates a proof of its correctness. The verifier's job is to verify the proof without redoing the entire analysis. Because of this feature, the prover only requires access to a subset of the data, such as a subset of nodes in a Merkle tree, rather than the entire dataset. This shift is significant because it offers a practical way to lower the costs of employing proofs of inclusion in smart contracts, particularly when massive datasets are involved.

\subsubsection{Applications} To prove the data is indeed stored in the blockchain, Herodotus \cite{herodotus} gives a new method called Storage Proofs to enable on-chain data access. It is essentially an on-chain accumulator that uses cryptographic means to improve access and verification of historical data on the Ethereum blockchain. The solution utilizes StarkWare's STARK proofs \cite{ben2018scalable}. This allows users to validate data from any point on the Ethereum blockchain without a third party. It combines proofs of inclusion for verifying the existence of data and proofs of computation for proving the execution of multi-step workflows. These proofs are essential for verifying the integrity and correctness of one or more components of a large data set, thus ensuring that the data has not only been stored but also remains untampered with and accurate over time.

Filecoin \cite{filecoin} has deployed a large-scale zkSNARK network that uses unused hard drive space from users worldwide to store files. Using zkSNARKs to generate proofs results in small proofs, and the verification process is swift and thus cheap. In some deployed constructions, zkSNARKs can reduce proof artifacts from much larger alternatives to very small constant-size proofs, substantially lowering verification overhead and making blockchain storage both more scalable and accessible.

\subsection{Smart Contract/Transaction Privacy}

\subsubsection{Motivation and Definition} While creating entirely new blockchain networks or rollups serves as a solution for privacy demands, there are also ZKP-based approaches which offer selective privacy for certain decentralized applications (dApps) and transactions on existing, transparent blockchains such as Ethereum. These applications could be utilized in cases where users want to make certain actions or information private, but continue to use an existing blockchain network. These applications exist on the smart contract execution layer of a blockchain, where dApps are designed to utilize ZKP to enable private transactions in certain contexts, such as transaction mixing, for example. This application domain can be defined as Smart Contract/Transaction Privacy, which mostly benefits from the privacy prong of ZK.

\subsubsection{Methodology} The core methodology in ZKP smart contract privacy centers around on-chain proof verification within a smart contract framework. Proof generation typically occurs off-chain due to its computationally intensive nature and the need for privacy in computation or the processing of sensitive data. Once a proof is generated, it is submitted to the blockchain. The contract assesses the proof against the protocol's predefined rules, and then acts accordingly if deemed valid, which could involve updating the blockchain's state, executing transactions, or any other protocol-specific actions. In practice, proof-system choice is driven by on-chain verification cost: Groth16 is widely used in deployed privacy contracts in part because its constant-size proofs can yield relatively low on-chain verification cost compared with larger-proof alternatives, whereas STARK proofs, though transparent and post-quantum, are typically larger and therefore often less attractive in gas-sensitive settings. Universal-setup schemes such as PLONK offer a middle ground--updatable structured reference strings and competitive proof sizes--and are increasingly adopted by projects that need circuit flexibility without per-application trusted setups.

\subsubsection{Applications}
\label{sec:blockchainPrivacyApps}

Tornado Cash \cite{tcash} employs zero-knowledge proofs for enabling transaction privacy and mixing on Ethereum. \textit{Statement proved:} set membership in a Merkle tree of deposit commitments, combined with a nullifier check that prevents double-spending. \textit{Privacy goal:} unlinking deposit and withdrawal addresses so that an observer cannot determine which deposit funds a given withdrawal. \textit{Engineering bottleneck:} in Tornado Cash's design, the effective anonymity set grows only as fast as new deposits arrive, and in typical implementations, on-chain verification for a withdrawal proof is on the order of a few hundred thousand gas. \textit{Limitations:} timing correlation between deposits and withdrawals can degrade anonymity in practice, and the fixed-denomination design (e.g., 0.1, 1, 10, 100 ETH) introduces a UX constraint that limits flexible-amount transfers.

Tornado Cash's deployment demonstrated that protocol-level unlinkability can attract both legitimate privacy demand and illicit use, leading to regulatory sanctions in several jurisdictions. This has given rise to the research problem of regulatory-compliant privacy solutions, which has been implemented in the Privacy Pools project, as delineated in \cite{buterin2023}. Privacy Pools extends the usage of zkSNARKS in other privacy solutions like Zcash and Tornado Cash. Rather than simply generating a ZKP to prove that a withdrawal attempt is linked to a specific deposit, users prove membership in a specific association set, which is a collection of transaction references, represented as a Merkle tree, from which a user's funds could have originated. Users define their set by providing the Merkle root as a public input. When withdrawing funds, the user does not directly prove a link to a specific transaction. Instead, they utilize zkSNARKS to generate proofs that validate their funds' origin from within this predefined, public set of transactions. By proving membership in a curated association set, users can demonstrate that their funds do not originate from flagged addresses, aligning transaction privacy with compliance requirements without sacrificing the cryptographic unlinkability of individual transfers.

More broadly, the tension between privacy and accountability has spurred several complementary design patterns. \emph{Selective disclosure} mechanisms allow authorized auditors to decrypt transaction metadata under predefined legal conditions while preserving unlinkability for ordinary observers~\cite{wust2022platypus}. \emph{Regulated set membership proofs} extend the association-set idea by requiring withdrawals to satisfy compliance predicates--e.g., that funds have not passed through sanctioned addresses within a configurable lookback window~\cite{buterin2023}. \emph{Accountability-focused designs} incorporate revocation or tracing capabilities that can be activated only through multi-party authorization, preventing unilateral surveillance while enabling lawful investigation~\cite{garman2016accountable}. These approaches remain at varying levels of maturity, and their interaction with jurisdiction-specific regulations is an active area of research.

Penumbra \cite{penumbra} is a fully private proof-of-stake network and decentralized exchange within the Cosmos ecosystem, offering a unique approach to transaction privacy and cross-blockchain interoperability using zero-knowledge proofs. Penumbra connects to the Cosmos blockchain ecosystem through IBC (the inter-blockchain communication protocol) and maintains all value in a multi-asset shielded pool, inspired by the Zcash Sapling design. This allows for private transactions in any IBC-compatible asset. All transactions on Penumbra are private by default, enabled by zkSNARKS which validate the correctness and legitimacy of transactions but shield the sender, receiver, and amount. Penumbra's decentralized exchange, called ZSwap, supports sealed-bid batch auctions and concentrated liquidity similar to Uniswap v3. This architecture prevents frontrunning and ensures that only the net flow across a pair of assets is revealed in each block. For cryptographic primitives, Penumbra utilizes BLS12-377 as the proving curve, which is compatible with the Groth16 proving system used. Penumbra may change to PLONK in the future.

A novel privacy problem exists within the space of on-chain decentralized autonomous organizations (DAOs), which are blockchain-based governance systems powered by smart contracts. Because of the transparent nature of public blockchains, DAO treasuries are completely public, which is a big issue for certain types of auctions. In a blog and research project by Griffin Dunaif and Dan Boneh, the authors design a system for a private DAO protocol utilizing ZKP \cite{pdao}. The protocol utilizes a master contract deployed on the Ethereum network to manage multiple DAOs. This contract allows anyone to send funds to a DAO, but only the DAO manager can withdraw them.
The life cycle of a DAO in this system comprises three steps: creation, deposit, and withdrawal. During DAO creation, the manager establishes the DAO without any on-chain transactions, and posts a Schnorr public key on the DAO website. To contribute funds to the DAO, users compute the Merkle tree leaf using the DAO public key, in which the deposit is recorded in the master contract. The DAO manager is able to privately monitor deposits and keep track of the DAO treasury using their secret key. When withdrawing funds, the DAO manager again uses this secret key within a SNARK proof to show that a specific batch of deposit leaves in the Merkle tree belong to the DAO, without publicly revealing the secret key. After verifying the proof, the master contract releases the withdrawal amount to the DAO manager.

We have seen niche-specific privacy dApps, but there also exist entirely private smart contract frameworks for private blockchain applications. One example is Hawk \cite{hawk}, which uses off-chain computation to conceal private portions of a smart contract. In Hawk, the programmer writes a smart contract with defined public and private components. The Hawk compiler will subsequently split the computation into pieces, where the private portion of the contract $\phi$-priv is executed off-chain and handles sensitive data and computations. This off-chain execution is managed by a trusted party, known as the manager, who can see the users' inputs and is expected to not disclose them. The manager's role is to perform the private computations and generate a ZK-SNARK attesting to the correctness of these computations without revealing any sensitive data inputs. The proofs are then verified on-chain, ensuring the integrity of the private computations while keeping them hidden from the public blockchain. The public portion of the contract $\phi$-pub, which executes on the blockchain, handles non-sensitive operations and provides transparency where necessary, but it does not deal directly with private data or currency transactions.

Mina, the previously mentioned Layer-1 blockchain, also provides a privacy-preserving smart contract framework called zkApps \cite{zkapp}. Mina zkApps are developed using the o1js TypeScript library and the Mina zkApp CLI, and comprise of two main components: a smart contract written with o1js, and a user interface (UI) for interaction. Upon zkApp interaction, the smart contract's code is executed locally in the user's web browser, where it generates a ZK-SNARK proof. This setup allows users to input data into the zkApp, which could be either private or public. Private data remains unseen by the blockchain, while public data may be stored on-chain or off-chain, depending on the zkApp's design. The prover function within the smart contract generates the SNARK proofs, maintaining user privacy by ensuring that sensitive data is processed locally and not disclosed on the blockchain. Once a user decides to submit a transaction to the chain, the transaction, containing the ZKP and associated state updates, is sent to the Mina network. The network verifies that the proof meets all constraints defined in the prover function. The state of a zkApp can be either on-chain or off-chain. On-chain state is stored directly on the Mina blockchain and offers limited storage space, while off-chain state refers to larger data stored elsewhere, like in external storage systems such as IPFS. In scenarios where off-chain storage is used, the zkApp updates an on-chain Merkle tree root of some fully off-chain Merkle tree. This method ensures that the integrity of both the proof and the associated account updates is maintained, allowing the Mina network to confirm the validity of the zkApp transactions and state changes.

\subsection{Blockchain-Based Proof of Identity}

\subsubsection{Motivation and Definition} Because of the aforementioned qualities of ZKP, there exists the unique use-case of a user proving group membership or identity without revealing any sensitive information about their identity, thus marrying the imperatives of authentication and privacy. While this can be implemented outside of the blockchain context (as will be covered in the ``Proof of Identity" section below), there exist many identity-proving applications using blockchain smart contracts and execution layers. This application domain can be defined as Blockchain-Based Proof of Identity, which takes advantage of privacy-preserving ZKP.

\subsubsection{Methodology} The core methodology behind blockchain-based proof of identity systems is the use of zero-knowledge proofs to enable users to prove membership in certain identity groups or ownership of credentials without revealing the actual identity information. This is achieved through cryptographic commitments to a user's information while keeping it secret. A privacy-preserving ZK identity proof can be easily verified by an on-chain smart contract, and this can serve as a private credential system secured by the blockchain, enabling decentralized applications without exposing sensitive information.

\subsubsection{Applications}

Semaphore is a framework for zero-knowledge signaling on Ethereum that allows users to broadcast support for an arbitrary string, without revealing their identity to anyone besides being approved to do so \cite{gurkan2020semaphore}. It uses Pedersen commitments to hide user identities in an incremental Merkle tree stored on-chain. Users generate zk-SNARK proofs showing that:
\begin{itemize}
    \item Their Pedersen commitment identity is valid, by proving it exists as a leaf in the incremental Merkle tree using the Merkle path.
    \item They know the secrets behind the Pedersen commitment.
    \item The unique, pseudorandomly derived nullifier has not been used before, preventing double-spending.
    \item The signal is properly authorized using an EdDSA signature verification within the circuit.
\end{itemize}
A smart contract handles logic and state management, such as adding identity commitments to the Merkle tree, updating the nullifier map, and adding successful signals to the signal map. This enables fully on-chain privacy applications with proof verification, such as anonymous voting and reputation systems \cite{gurkan2020semaphore}.

Semaphore can serve as a verifiable credential protocol at the base level, as it allows developers to create identities, identity groups, and use identity commitments to prove group membership. An example of a prominent project that utilizes Semaphore for verifiable credentials on Ethereum is World ID, which is the proof-of-personhood verification system for the WorldCoin protocol \cite{worldcoin}. When a user creates a unique World ID from their biometric data, their ID is enrolled in a group of verified World ID users. World ID's can then be safely verified without revealing the World ID public key.

The Galxe protocol is a self-sovereign identity service centered around verifiable credentials \cite{galxe}. It addresses the digital identity multiplicity problem by embedding numerous identity commitments into a single user credential, enabling holders to connect identities across platforms privately. Credential holders can use ZKP to selectively prove requisite information of their identity, while maintaining a pseudonym. Like Semaphore, Galxe constructs identity commitments by computing the Poseidon hash of the private secret identity and a private internal nullifier. This can then be distinguished in a zero-knowledge proof, which is verified by an on-chain smart contract and attests to the user's identity. The internal and external nullifiers generate deterministic nullifiers per verification to prevent double-spending. The current verification stack of the Galxe protocol is BabyZK, which uses the BN254 curve, Groth16 proofs and Poseidon commitments. Use cases of the Galxe protocol could include sybil-resistant reputation systems, access control, achievement aggregation, and personal data markets with privacy.

Aleo, a Layer 1 privacy blockchain mentioned in this paper above, includes its own ZK identity verification protocol called zPass \cite{zpass}. zPass is a straightforward implementation due to the nature of enshrined zero-knowledge proofs and private execution available in Aleo. However, zPass also enables users to obtain anonymous credentials from existing identity documents, facilitating real-world adoption without necessitating changes to the protocol. Like Galxe protocol, it allows for selective attribute disclosure, enabling users to prove identity assertions across multiple credentials.

\subsection{Supply Chain/Enterprise Blockchain Privacy}

\subsubsection{Motivation and Definition}
Supply chains and global enterprises are increasingly adopting blockchain technology to improve transparency, provenance, immutability, and accountability. However, this shift towards blockchain adoption in supply chain comes with significant privacy concerns. In a typical supply chain, stakeholders, including suppliers, manufacturers, distributors, and retailers, share sensitive data such as pricing, inventory levels, and production schedules. The public and immutable nature of open blockchains could expose proprietary or sensitive information to competitors or the public. This trade-off between the need for transparency and privacy requirements has led to the exploration of privacy-enhancing technologies within blockchain frameworks for supply chain and enterprise applications.

\subsubsection{Methodology}
Implementing privacy in supply chain and enterprise blockchains involves several strategic approaches. The use of permissioned blockchain architectures creates a controlled ecosystem where only verified participants can access data, ensuring that sensitive information is shared exclusively with trusted stakeholders. Additionally, zero-knowledge proofs (ZKPs) allow for the verification of product authenticity and compliance without disclosing detailed process histories, thereby maintaining consumer confidence in product safety and quality. To address the complex nature of modern supply chains, data segmentation and encryption techniques are employed. By storing only cryptographic hashes on-chain and keeping detailed records off-chain or encrypted, these methods protect business-sensitive information and reduce the risk of exposing trade secrets. Smart contracts further enhance privacy by automating compliance and access control based on predefined rules, ensuring that data disclosure is governed by necessity and consent. These measures collectively protect consumer interests and maintain trust among supply chain participants.

\subsubsection{Applications}
qedit \cite{QEDIT} provides privacy-enhancing technology for enterprise blockchains, enabling secure collaboration and data sharing among parties without revealing sensitive information. It utilizes ZKPs to ensure that transactions are valid while keeping the transaction content private. It is designed for enterprises looking to monetize data assets safely, enhance business analytics, and derive actionable insights in a secure manner. The platform is cloud-hosted, highly scalable, and integrates easily with existing database systems for quick deployment. qedit features a configurable dashboard, advanced reporting, and real-time notifications to provide businesses with critical intelligence efficiently. It is aimed at transforming the way companies collaborate on sensitive data, ensuring privacy while enabling data-driven decisions.

zk-BeSC \cite{nasri2023zkbesc} introduces a blockchain-based framework for supply chain management that utilizes polynomial ZKPs to ensure privacy during transactions. The framework is designed to enable confidential transactions among supply chain participants, preserving the privacy of sensitive data while maintaining the traceability and immutability features of blockchain technology. It leverages advanced cryptographic techniques, including homomorphic encryption and elliptic curve pairings, to prove knowledge of polynomials without disclosing them. Implemented on the Ethereum testnet with a web3 application, zk-BeSC demonstrates efficient proof performance and reduced gas consumption for verification, addressing key privacy concerns in supply chain management.

Sahai \textit{et al.} \cite{sahai2020enabling} present a blockchain-based solution for improving privacy and traceability in supply chains. This approach leverages Hyperledger Fabric and employs cryptographic tools like ZKPs to protect sensitive business data while ensuring the ability to trace product provenance and contamination. The model supports operations such as product entry, exit, transfer, merge, split, and processing within the supply chain, enabling efficient and private tracing of products from origin to consumer.

Beyond provenance tracking, three additional enterprise privacy patterns have emerged. \textit{Private compliance proofs} allow a regulated entity to demonstrate that a dataset satisfies a policy predicate--e.g., that all transactions fall below a reporting threshold--without revealing the underlying records~\cite{narula2018zkledger}. \textit{B2B attestations} enable one firm to prove aggregate properties of an invoice batch or shipment manifest (total value, item count, origin country) to a counterparty without disclosing line-item detail~\cite{nasri2023zkbesc}. \textit{Selective disclosure credentials} let an organization extract verifiable assertions from a signed credential bundle--for instance, proving ISO-9001 certification status without revealing the full audit report~\cite{zk-creds}. A shared bottleneck across these patterns is that circuit complexity scales with schema richness: the more fields a compliance predicate or credential format contains, the larger the constraint system and the longer the proving time.

\begin{table}[htbp]
\caption{Enterprise ZKP patterns: statement proved, primary bottleneck, and deployment maturity.}
\label{tab:enterprise_patterns}
\centering
\resizebox{0.6\textwidth}{!}{%
\begin{tabular}{|l|l|l|l|l|}
\hline
\textbf{Pattern} & \textbf{Statement proved} & \textbf{Bottleneck} & \textbf{Maturity}  \\
\hline
Provenance tracking & Product origin \& chain of custody & Hash-chain depth & Pilot  \\
\hline
Private compliance & Policy predicate over private data & Predicate circuit size & Research  \\
\hline
B2B attestation & Aggregate properties of records & Schema-dependent circuit & Proof-of-concept  \\
\hline
Selective disclosure & Assertion from signed credential & Signature verification in-circuit & Pilot \\
\hline
\end{tabular}}
\end{table}

\subsection{Proof of Reserves}
\subsubsection{Motivation and Definition}
It has always been important in financial markets for companies to demonstrate their reserve assets to prove their solvency to savers. The global financial system often operates in an undercollateralized and highly opaque manner, relying heavily on trust in a central entity (either the system itself or a third-party certifier), but this can create fraud risks, mismanagement, and privacy breaches. Zero-knowledge proof of Reserves (ZK-PoR) (such as \cite{conley2023instant}) provides a trustless mechanism to verify reserves, enhancing trust and security in decentralized financial systems without sacrificing privacy. This innovation is particularly important for cryptocurrency exchanges and wallets, as users need verifiable assurance that their assets are being held securely without exposing those assets to potential threats. Specific to the scenario of a virtual currency exchange, the proof generated by this method can ensure that the verifier obtains proof of the exchange's repayment ability without knowing the specific amount of the exchange's reserves, the identity of the individual account holder, or any transaction details. Proof, thereby ensuring transparency on reserve adequacy while maintaining privacy and security.
\subsubsection{Methodology}
The methodology behind ZK-PoR involves several key steps to ensure secure and private verification of assets. Initially, the entity holding the reserve constructs a commitment to the asset without revealing its details, using cryptography to generate proof of possession. The proof is rooted in a zero-knowledge proof algorithm and is then transmitted to the verifier. Validators use the same cryptographic algorithm to verify proofs without knowing the actual reserves, the identities of asset holders, or transaction details. The process often employs complex mathematical structures such as homomorphic encryption, elliptic curve encryption, and Merkle trees to ensure the integrity and confidentiality of the proof. The trust established by this approach stems from mathematical proof rather than the reputation or authority of the entity.
\subsubsection{Applications}
ZK-PoR has numerous applications in the fields of blockchain and financial technology, especially in enhancing privacy and trust. In the financial domain, particularly within cryptocurrency exchanges, proving solvency without compromising sensitive information confidentiality is a critical concern.

Provisions \cite{dagher2015provisions} addresses this challenge using ZK-PoR. This technology enables a Bitcoin exchange, or any cryptocurrency platform, to transparently demonstrate that it possesses sufficient funds to cover all its obligations to users without disclosing the exact amounts held or the identities of the account holders. Allowing an exchange to prove its liquidity addresses the common concern of potential insolvency. This approach also reduces the risk of exposing sensitive financial data to privacy breaches or malicious exploitation.

Another application, Proven\cite{proven}, leverages ZK-PoR to provide a decentralized platform that enables companies and financial institutions to verifiably prove their liquidity or asset reserves without revealing specific asset values or compromising the confidentiality of their operations. These applications highlight ZK-PoR's versatility in solving the twin challenges of transparency and privacy in digital finance, providing exchanges and institutions with a powerful solution to build trust with users and regulators while protecting sensitive financial data.

Despite its promise, ZK-PoR carries several under-examined assumptions and failure modes. First, \textit{completeness of liabilities} is not guaranteed: a custodian can exclude accounts from the liability Merkle tree, understating obligations while the proof remains mathematically valid. Second, proofs are \textit{point-in-time snapshots}--assets can be moved immediately after proof generation, so a passing proof does not imply ongoing solvency. Third, the \textit{address-to-reserve mapping} is self-attested by the custodian; without an independent on-chain audit linking addresses to the proving entity, the proof may cover borrowed or temporarily parked assets. Finally, there exists an inherent \textit{auditability--privacy tradeoff}: stronger privacy protections (e.g., hiding individual account balances) reduce the ability of external parties to detect account omissions, making it harder to catch the very fraud the mechanism is designed to prevent.

\section{Non-Blockchain Applications} \label{sec:nonblockchain}

In this section, we begin with ML/AI as the main non-blockchain focus, then group the remaining application topics into a single category with a summary table.

\begin{table}[htbp]
\centering
\caption{Overview of non-blockchain applications of Zero-Knowledge Proofs}
\resizebox{0.7\textwidth}{!}{%
\begin{tabularx}{\textwidth}{|l|X|}
\hline
\textbf{Use Case} & \textbf{Representative works/projects} \\
\hline
ML/AI & zkCNN, vCNN, zkDL, Kaizen, zkLLM, ezkl, Modulus, Giza, TensorPlonk \\
\hline
Other applications & Proofs of Identity, zk-creds; voting ZK arguments, PhotoProof, VeeDo, Collaborative zkSNARKS, ZKP2P \\
\hline
\end{tabularx}}
\label{tab:nonblockchain_use_case_projects}
\end{table}

\subsection{Machine Learning and AI}

\subsubsection{Motivation and Definition}
In a world where AI-generated material increasingly mimics human-created information, the potential use of zero-knowledge cryptography could assist us in determining that a specific piece of content was produced by applying a specific model to a given input. If a zero-knowledge circuit representation is built for them, this could give a technique for verifying outputs from large language models like GPT-4 \cite{achiam2023gpt}, text-to-image models, or any other models. The zero-knowledge quality of these proofs allows us to hide sections of the input or the model if necessary. A good example is enabling users to view the model's inference results without knowing the details of the model and prove that this result really comes from a specific model and input.
Zero-knowledge machine learning (ZKML) is a means to protect data privacy during model training and inference. They enable a data owner or a model owner to demonstrate the accuracy of a computation (such as the prediction of a machine learning model) without exposing any information about the data or the computation itself. This is especially effective in circumstances involving sensitive data.

\subsubsection{Methodology}
In a typical machine learning situation, an application service provider (the prover) wants to provide a machine learning model it owns to the user (the verifier) while keeping the model private. Provers can use zero-knowledge proofs to show that they have indeed performed a computation using a particular model without exposing the model or the computation process itself.

\subsubsection{Applications}
ZKPs are commonly employed in machine learning, notably privacy-preserving and federated learning. service providers (provers) can use ZKP to prove the correctness of their model predictions without revealing the model itself to prevent model theft. Current zero-knowledge systems, even with strong hardware acceleration, remain far from efficiently proving inference or training at the scale of today's largest language models (LLMs), but there has been considerable progress in establishing proofs of smaller models.
In verifying machine learning model predictions, vCNN \cite{lee2020vcnn} uses commit-and-prove to combine the typical quadratic arithmetic program (QAP) with the polynomial QAP in pairing-based zero-knowledge proofs. It supports convolutional neural networks and validates them using polynomial multiplications.
However, by presenting a novel sumcheck technique, zkCNN \cite{liu2021zkcnn} could prove fast Fourier transformations and convolutions in a linear prover time. In the authors' evaluated setting, zkCNN reports substantially faster proving than vCNN, including a reported 34$\times$ improvement for the studied convolution workload.
zkDL\cite{sun2023zkdl} addresses the problem of zero-knowledge proofs for deep learning training. The main challenge it addresses is verifying the nonlinear computations inherent in neural networks, especially the ReLU activation function and its backpropagation. By introducing zkReLU, zkDL can efficiently handle these non-arithmetic operations without resorting to polynomial approximations, which are usually computationally expensive and less accurate. The FAC4DNN utilized by zkDL is an arithmetic circuit that aggregates proofs from different layers and training steps. This design bypasses traditional sequential proof generation and greatly reduces computational and communication overheads. zkDL reports compatibility with tensor-based workloads and, in the authors' evaluation, sub-second proofs per batch update for networks with up to 10 million parameters.

The research by Sanjam Garg et al.\cite{garg2023experimenting} explores the practical application of zero-knowledge proofs in verifying the training of machine learning models. It focuses on experimental settings to verify the feasibility of such proofs, taking into account computational complexity and scalability issues. Their work emphasizes the need to balance strong security guarantees with practical overhead, ensuring that the proofs are concise and the prover's workload is manageable. By experimenting with various configurations and optimization techniques, it provides insights into making zero-knowledge proofs applicable to real-world deep learning applications, which is very suitable as a reference.
Another recent study, Kaizen\cite{abbaszadeh2024zero}, proposed a zero-knowledge proof system designed for deep neural networks (DNNs). It ensures that the submitted model is correctly trained on the submitted dataset without leaking any other information. Kaizen adopts a sum-check-based proof system optimized for the gradient descent algorithm and recursively combines these proofs in multiple iterations. This recursive combination is designed so that proof size and verifier time do not grow linearly with the number of iterations, improving scalability. Kaizen has the ability to handle large models such as VGG-11, and is more practical than general recursive proofs, significantly reducing prover runtime and memory overhead.

In the study of the reasoning process, zkLLM\cite{sun2024zkllm} focuses on providing zero-knowledge proofs for large language models, ensuring that the reasoning results are verifiable without revealing the underlying model or data. This is especially important for applications that require strong privacy protection, such as medical or financial fields where data confidentiality is critical. zkLLM uses advanced encryption technology to efficiently generate and verify proofs, and maintains the privacy of model parameters and input data through cryptographic means, enhancing trust in deployed AI systems.
In addition to the issue of the trustworthiness of its predictions, the model's reliance on opaque data sources has become a new challenge. People frequently desire to keep the inputs and parameters of machine learning models private and unknown to the general public. Because the input data may contain sensitive information such as personal finances or biometrics, and the model parameters may also hold critical secrets. To specify provers and verifiers, many ZKML tools represented by the ezkl library \cite{ezkl} can implement higher-level descriptions of machine learning models or other computational graph programs. A prover can demonstrate that a particular output is produced by running a specific neural network on a specific data set.

Modulus Labs \cite{modulus} is also developing a zero-knowledge machine learning solution, enabling trustful integration of AI outputs into blockchain systems without revealing sensitive data or model details. This allows developers to retain ownership and control over their AI models rather than relying on centralized platforms to host and manage their models. And their new zero-knowledge prover, Remainder\cite{ScalingIntelligence}, is a fast AI prover for AI inference. It is based on a verifiable decision forest inference circuit using GKR protocol\cite{goldwasser2015delegating}.
Giza \cite{Giza} is a machine learning platform built on StarkNet that can be used to deploy and extend machine learning models, as well as solve the interoperability issues faced in the use of cloud-based machine learning models, performance, and transparency issues. It uses the ONNX open format to improve interoperability. Through a series of common operators and file formats it defines, developers can freely use TensorFlow, PyTorch, Scikit-Learn, and other frameworks and tools. Since StarkNet runs on the Layer2 network, it can enable any decentralized application to achieve unlimited computing scale without affecting the composability and security of Ethereum. This design reduces some infrastructure burdens on developers, allowing greater focus on model development. Another benefit of being based on StarkNet is that most functions are managed by the blockchain, which makes it easier to monitor, track, and manage the model, greatly improving transparency.

\subsection{Other Non-Blockchain Applications}
\label{sec:other-nonblockchain}

\begin{table}[htbp]
\centering

\caption{Summary of other non-blockchain applications of ZKPs}
\resizebox{0.8\textwidth}{!}{%
\begin{tabularx}{\textwidth}{|l|X|X|}
\hline
\textbf{Area} & \textbf{What is proved} & \textbf{Representative works/projects} \\
\hline
Identity and credentials & Membership or attribute validity without disclosure & Feige--Fiat--Shamir identification \cite{fiege1987zero}; zk-creds \cite{zk-creds} \\
\hline
Voting & Ballot validity and tally correctness with privacy & Non-interactive ZK arguments for voting \cite{groth2005non} \\
\hline
Media authentication & Valid transform of an image under allowed edits & PhotoProof \cite{naveh2016photoproof} \\
\hline
Randomness and timelocks & Correctness of delay functions and derived randomness & VeeDo \cite{veedo2023} \\
\hline
Collaborative computation & Correct joint computation over private inputs & Collaborative zkSNARKS \cite{ozdemir2022experimenting} \\
\hline
Web2--Web3 payment bridging & Validity of off-chain payment evidence without revealing contents & ZKP2P \cite{zkp2p} \\
\hline
\end{tabularx}}
\label{tab:other_apps_table}
\end{table}

Proof of identity protocols use ZKPs to show possession of a credential, or membership in a set, without disclosing the credential itself. Early work formalized interactive identification based on proving knowledge of a secret \cite{fiege1987zero}. More recent systems such as zk-creds turn existing identity documents into anonymous credentials that can be verified without issuers maintaining signing keys \cite{zk-creds}.
The remaining applications in Table~\ref{tab:other_apps_table} follow the same theme: prove that a rule was followed, without exposing the underlying data. PhotoProof targets image integrity under allowed edits \cite{naveh2016photoproof}. Voting arguments target ballot correctness while keeping votes private \cite{groth2005non}. VeeDo uses ZK machinery around delay functions for timelocks and randomness \cite{veedo2023}. Collaborative zkSNARKS support joint analytics without sharing inputs \cite{ozdemir2022experimenting}, and ZKP2P proves properties of payment emails (e.g., DKIM signatures) to bridge payment rails and on-chain settlement without revealing email contents \cite{zkp2p}.

\section{Conclusion and Future Work} \label{sec:conclusion}
Zero-knowledge proofs have matured from theoretical constructs into practical tools that bridge cryptographic research and real-world deployment. This survey examined their applications across blockchain privacy, scalable computation, identity, and emerging non-blockchain domains such as verifiable machine learning and regulatory compliance. Despite rapid progress, significant challenges remain in proof-system optimization, developer tooling, and broader adoption beyond the blockchain ecosystem.
An especially promising intersection lies in merging ZKPs with game-theoretic mechanisms--for instance, enabling truthful auction bidding without revealing sensitive information about underlying assets, akin to the private DAO applications discussed in Section~\ref{sec:blockchainPrivacyApps}. Privacy-preserving financial price discovery, private order-book exchanges, dark pools using the AMM model, and encrypted transaction mempools for mitigating maximal extractable value (MEV) all represent open research frontiers.

{We see several directions for future work in this area}:

\textbf{Standardization and interoperability:}  Efforts such as ZKProof.org have begun defining shared terminology and security guidelines, yet the ecosystem still lacks common intermediate representations and universal proof formats. Establishing interoperability standards across proof systems, together with reproducible benchmark suites, would lower integration costs and accelerate adoption by practitioners who currently face vendor lock-in.

\textbf{Post-quantum considerations:}  Transparent-setup systems such as STARKs and FRI-based protocols already avoid pairing-based assumptions vulnerable to quantum adversaries. Concurrently, lattice- and hash-based SNARKs are under active investigation. A key open question is how to balance quantum resistance against proof size and on-chain verification cost, since post-quantum constructions currently incur substantial overhead.

\textbf{Circuit security and verification tooling:}  Under-constrained circuits and hand-written bugs remain a significant practical risk, as discussed in the DSL landscape of Section~\ref{sec:DSLs}. Formal verification of constraint systems, static-analysis tooling for ZK-specific DSLs, and automated under-constraint detection are critical for closing the gap between theoretical soundness guarantees and deployed system security.

\textbf{Performance roadmap:}  Proof recursion and composition, folding schemes such as Nova and incremental verifiable computation (IVC), and hardware acceleration via FPGAs and ASICs (Section~3.4) together chart a path toward substantially lower-latency proving for latency-sensitive applications. Reducing prover costs by orders of magnitude would unlock use cases--such as universal synchronous composability among Layer-2 rollups--that are currently bottlenecked by proof generation time.

\section*{Acknowledgments}
This material is based in part on work supported by AFOSR under award number FA9550-23-1-0312. Any opinions, findings, and conclusions, or recommendations expressed in this material are those of the author(s) and do not necessarily reflect the views of any funding agencies. We thank Prof Yupeng Zhang for his helpful comments on the paper.

%Bibliography
\bibliographystyle{unsrturl}
\bibliography{bibliography}

@article{sun2021survey,
  author    = {Sun, Xiaoqiang and Yu, F. Richard and Zhang, Peng and Sun, Zhiwei and Xie, Weixin and Peng, Xiang},
  title     = {A Survey on Zero-Knowledge Proof in Blockchain},
  journal   = {IEEE Network},
  volume    = {35},
  number    = {4},
  pages     = {198--205},
  year      = {2021},
  publisher = {IEEE},
  doi       = {10.1109/MNET.011.2000473},
  url       = {https://doi.org/10.1109/MNET.011.2000473}
}

@article{morais2019survey,
  author    = {Morais, Eduardo and Koens, Tommy and Van Wijk, Cees and Koren, Aleksei},
  title     = {A Survey on Zero-Knowledge Range Proofs and Applications},
  journal   = {{SN} Applied Sciences},
  volume    = {1},
  pages     = {946},
  year      = {2019},
  publisher = {Springer},
  doi       = {10.1007/s42452-019-0989-z},
  url       = {https://doi.org/10.1007/s42452-019-0989-z}
}

@misc{lee2020vcnn,
  author       = {Lee, Seunghwa and Ko, Hankyung and Kim, Jihye and Oh, Hyunok},
  title        = {{vCNN}: Verifiable Convolutional Neural Network Based on {zk-SNARKs}},
  howpublished = {{IACR} Cryptology ePrint Archive, Report 2020/584},
  year         = {2020},
  url          = {https://eprint.iacr.org/2020/584}
}

@article{petkus2019and,
  author  = {Petkus, Maksym},
  title   = {Why and How {zk-SNARK} Works},
  journal = {arXiv preprint arXiv:1906.07221},
  year    = {2019},
  url     = {https://arxiv.org/abs/1906.07221}
}

@article{zhang2019security,
  author    = {Zhang, Rui and Xue, Rui and Liu, Ling},
  title     = {Security and Privacy on Blockchain},
  journal   = {ACM Computing Surveys},
  volume    = {52},
  number    = {3},
  articleno = {51},
  numpages  = {34},
  year      = {2019},
  publisher = {Association for Computing Machinery},
  doi       = {10.1145/3316481},
  url       = {https://doi.org/10.1145/3316481}
}

@article{o2019design,
  author    = {O'Donoghue, Odhran and Vazirani, Anuraag A. and Brindley, David and Meinert, Edward},
  title     = {Design Choices and Trade-Offs in Health Care Blockchain Implementations: Systematic Review},
  journal   = {Journal of Medical Internet Research},
  volume    = {21},
  number    = {5},
  pages     = {e12426},
  year      = {2019},
  publisher = {{JMIR} Publications},
  doi       = {10.2196/12426},
  url       = {https://doi.org/10.2196/12426}
}

@article{achiam2023gpt,
  author  = {Achiam, Josh and Adler, Steven and Agarwal, Sandhini and Ahmad, Lama and Akkaya, Ilge and Aleman, Florencia Leoni and Almeida, Diogo and Altenschmidt, Janko and Altman, Sam and Anadkat, Shyamal and others},
  title   = {{GPT}-4 Technical Report},
  journal = {arXiv preprint arXiv:2303.08774},
  year    = {2023},
  url     = {https://arxiv.org/abs/2303.08774}
}

@misc{conley2023instant,
  author       = {Conley, Trevor and Diaz, Nilsson and Espada, Diego and Kuruvilla, Alvin and Mayone, Stenton and Fu, Xiang},
  title        = {Instant Zero Knowledge Proof of Reserve},
  howpublished = {{IACR} Cryptology ePrint Archive, Report 2023/1156},
  year         = {2023},
  url          = {https://eprint.iacr.org/2023/1156}
}

@misc{ben2018scalable,
  author       = {Ben-Sasson, Eli and Bentov, Iddo and Horesh, Yinon and Riabzev, Michael},
  title        = {Scalable, Transparent, and Post-Quantum Secure Computational Integrity},
  howpublished = {{IACR} Cryptology ePrint Archive, Report 2018/046},
  year         = {2018},
  url          = {https://eprint.iacr.org/2018/046}
}

@misc{sun2023zkdl,
  author       = {Sun, Haochen and Bai, Tonghe and Li, Jason and Zhang, Hongyang},
  title        = {{ZkDL}: Efficient Zero-Knowledge Proofs of Deep Learning Training},
  howpublished = {{IACR} Cryptology ePrint Archive, Report 2023/1174},
  year         = {2023},
  url          = {https://eprint.iacr.org/2023/1174}
}

@misc{abbaszadeh2024zero,
  author       = {Abbaszadeh, Kasra and Pappas, Christodoulos and Papadopoulos, Dimitrios and Katz, Jonathan},
  title        = {Zero-Knowledge Proofs of Training for Deep Neural Networks},
  howpublished = {{IACR} Cryptology ePrint Archive, Report 2024/162},
  year         = {2024},
  url          = {https://eprint.iacr.org/2024/162}
}

@misc{sun2024zkllm,
  author       = {Sun, Haochen and Li, Jason and Zhang, Hongyang},
  title        = {{zkLLM}: Zero Knowledge Proofs for Large Language Models},
  howpublished = {arXiv preprint arXiv:2404.16109},
  year         = {2024},
  url          = {https://arxiv.org/abs/2404.16109}
}

@incollection{rivest1978data,
  author    = {Rivest, Ronald L. and Adleman, Leonard and Dertouzos, Michael L.},
  title     = {On Data Banks and Privacy Homomorphisms},
  booktitle = {Foundations of Secure Computation},
  editor    = {DeMillo, Richard A. and Dobkin, David P. and Jones, Anita K. and Lipton, Richard J.},
  pages     = {169--180},
  year      = {1978},
  publisher = {Academic Press}
}

@inproceedings{ben2013snarks,
  author    = {Ben-Sasson, Eli and Chiesa, Alessandro and Genkin, Daniel and Tromer, Eran and Virza, Madars},
  title     = {{SNARKs} for {C}: Verifying Program Executions Succinctly and in Zero Knowledge},
  booktitle = {Advances in Cryptology--CRYPTO 2013},
  series    = {Lecture Notes in Computer Science},
  volume    = {8042},
  pages     = {90--108},
  year      = {2013},
  publisher = {Springer},
  doi       = {10.1007/978-3-642-40084-1_6},
  url       = {https://doi.org/10.1007/978-3-642-40084-1_6}
}

@inproceedings{liu2021zkcnn,
  author    = {Liu, Tianyi and Xie, Xiang and Zhang, Yupeng},
  title     = {{zkCNN}: Zero Knowledge Proofs for Convolutional Neural Network Predictions and Accuracy},
  booktitle = {Proceedings of the 2021 ACM SIGSAC Conference on Computer and Communications Security},
  pages     = {2968--2985},
  year      = {2021},
  publisher = {Association for Computing Machinery},
  doi       = {10.1145/3460120.3485379},
  url       = {https://doi.org/10.1145/3460120.3485379}
}

@inproceedings{yao1982protocols,
  author    = {Yao, Andrew C.},
  title     = {Protocols for Secure Computations},
  booktitle = {23rd Annual Symposium on Foundations of Computer Science (FOCS 1982)},
  pages     = {160--164},
  year      = {1982},
  publisher = {IEEE},
  doi       = {10.1109/SFCS.1982.38},
  url       = {https://doi.org/10.1109/SFCS.1982.38}
}

@inproceedings{groth2005non,
  author    = {Groth, Jens},
  title     = {Non-Interactive Zero-Knowledge Arguments for Voting},
  booktitle = {Applied Cryptography and Network Security (ACNS 2005)},
  series    = {Lecture Notes in Computer Science},
  volume    = {3531},
  pages     = {467--482},
  year      = {2005},
  publisher = {Springer},
  doi       = {10.1007/11496137_32},
  url       = {https://doi.org/10.1007/11496137_32}
}

@inproceedings{bulletproofs,
  author    = {B{\"u}nz, Benedikt and Bootle, Jonathan and Boneh, Dan and Poelstra, Andrew and Wuille, Pieter and Maxwell, Gregory},
  title     = {Bulletproofs: Short Proofs for Confidential Transactions and More},
  booktitle = {2018 IEEE Symposium on Security and Privacy (SP)},
  pages     = {315--334},
  year      = {2018},
  publisher = {IEEE},
  doi       = {10.1109/SP.2018.00020},
  url       = {https://doi.org/10.1109/SP.2018.00020}
}

@inproceedings{groth16,
  author    = {Groth, Jens},
  title     = {On the Size of Pairing-Based Non-Interactive Arguments},
  booktitle = {Advances in Cryptology--EUROCRYPT 2016},
  series    = {Lecture Notes in Computer Science},
  volume    = {9666},
  pages     = {305--326},
  year      = {2016},
  publisher = {Springer},
  doi       = {10.1007/978-3-662-49896-5_11},
  url       = {https://doi.org/10.1007/978-3-662-49896-5_11}
}

@inproceedings{naveh2016photoproof,
  author    = {Naveh, Assa and Tromer, Eran},
  title     = {Photoproof: Cryptographic Image Authentication for Any Set of Permissible Transformations},
  booktitle = {2016 IEEE Symposium on Security and Privacy (SP)},
  pages     = {255--271},
  year      = {2016},
  publisher = {IEEE},
  doi       = {10.1109/SP.2016.23},
  url       = {https://doi.org/10.1109/SP.2016.23}
}

@inproceedings{ozdemir2022experimenting,
  author    = {Ozdemir, Alex and Boneh, Dan},
  title     = {Experimenting with Collaborative {zk-SNARKs}: Zero-Knowledge Proofs for Distributed Secrets},
  booktitle = {31st USENIX Security Symposium (USENIX Security 22)},
  pages     = {4291--4308},
  year      = {2022},
  publisher = {USENIX Association},
  url       = {https://www.usenix.org/conference/usenixsecurity22/presentation/ozdemir}
}

@inproceedings{goldwasser1985knowledge,
  author    = {Goldwasser, Shafi and Micali, Silvio and Rackoff, Charles},
  title     = {The Knowledge Complexity of Interactive Proof-Systems},
  booktitle = {Proceedings of the Seventeenth Annual ACM Symposium on Theory of Computing},
  pages     = {291--304},
  year      = {1985},
  publisher = {Association for Computing Machinery},
  doi       = {10.1145/22145.22178},
  url       = {https://doi.org/10.1145/22145.22178}
}

@inproceedings{hawk,
  author    = {Kosba, Ahmed and Miller, Andrew and Shi, Elaine and Wen, Zikai and Papamanthou, Charalampos},
  title     = {Hawk: The Blockchain Model of Cryptography and Privacy-Preserving Smart Contracts},
  booktitle = {2016 IEEE Symposium on Security and Privacy (SP)},
  pages     = {839--858},
  year      = {2016},
  publisher = {IEEE},
  doi       = {10.1109/SP.2016.55},
  url       = {https://doi.org/10.1109/SP.2016.55}
}

@inproceedings{kate2010constant,
  author    = {Kate, Aniket and Zaverucha, Gregory M. and Goldberg, Ian},
  title     = {Constant-Size Commitments to Polynomials and Their Applications},
  booktitle = {Advances in Cryptology--ASIACRYPT 2010},
  series    = {Lecture Notes in Computer Science},
  volume    = {6477},
  pages     = {177--194},
  year      = {2010},
  publisher = {Springer},
  doi       = {10.1007/978-3-642-17373-8_11},
  url       = {https://doi.org/10.1007/978-3-642-17373-8_11}
}

@inproceedings{fiat1986prove,
  author    = {Fiat, Amos and Shamir, Adi},
  title     = {How to Prove Yourself: Practical Solutions to Identification and Signature Problems},
  booktitle = {Advances in Cryptology--CRYPTO'86},
  series    = {Lecture Notes in Computer Science},
  volume    = {263},
  pages     = {186--194},
  year      = {1987},
  publisher = {Springer},
  doi       = {10.1007/3-540-47721-7_12},
  url       = {https://doi.org/10.1007/3-540-47721-7_12}
}

@inproceedings{gurkan2020semaphore,
  author = {Gurkan, Kobi and Koh, Wei Jie and Whitehat, Barry},
  title  = {Community Proposal: Semaphore: Zero-Knowledge Signaling on Ethereum},
  year   = {2020},
  month  = mar,
  day    = {31},
  url    = {https://docs.zkproof.org/pages/standards/accepted-workshop3/proposal-semaphore.pdf}
}

@inproceedings{dagher2015provisions,
  author    = {Dagher, Gaby G. and B{\"u}nz, Benedikt and Bonneau, Joseph and Clark, Jeremy and Boneh, Dan},
  title     = {Provisions: Privacy-Preserving Proofs of Solvency for Bitcoin Exchanges},
  booktitle = {Proceedings of the 22nd ACM SIGSAC Conference on Computer and Communications Security},
  pages     = {720--731},
  year      = {2015},
  publisher = {Association for Computing Machinery},
  doi       = {10.1145/2810103.2813674},
  url       = {https://doi.org/10.1145/2810103.2813674}
}

@inproceedings{nasri2023zkbesc,
  author    = {Nasri, Jaouaher Zouari and Rais, Helmi},
  title     = {{zk-BeSC}: Confidential Blockchain Enabled Supply Chain Based on Polynomial Zero-Knowledge Proofs},
  booktitle = {2023 International Wireless Communications and Mobile Computing (IWCMC)},
  pages     = {1472--1478},
  year      = {2023},
  publisher = {IEEE},
  doi       = {10.1109/IWCMC58020.2023.10183017},
  url       = {https://doi.org/10.1109/IWCMC58020.2023.10183017}
}

@inproceedings{sahai2020enabling,
  author    = {Sahai, Shubham and Singh, Nitin and Dayama, Pankaj},
  title     = {Enabling Privacy and Traceability in Supply Chains Using Blockchain and Zero Knowledge Proofs},
  booktitle = {2020 IEEE International Conference on Blockchain (Blockchain)},
  pages     = {134--143},
  year      = {2020},
  publisher = {IEEE},
  doi       = {10.1109/Blockchain50366.2020.00024},
  url       = {https://doi.org/10.1109/Blockchain50366.2020.00024}
}

@inproceedings{xie2022zkbridge,
  author    = {Xie, Tiancheng and Zhang, Jiaheng and Cheng, Zerui and Zhang, Fan and Zhang, Yupeng and Jia, Yongzheng and Boneh, Dan and Song, Dawn},
  title     = {{zkBridge}: Trustless Cross-Chain Bridges Made Practical},
  booktitle = {Proceedings of the 2022 ACM SIGSAC Conference on Computer and Communications Security},
  pages     = {3003--3017},
  year      = {2022},
  publisher = {Association for Computing Machinery},
  doi       = {10.1145/3548606.3560652},
  url       = {https://doi.org/10.1145/3548606.3560652}
}

@inproceedings{garg2023experimenting,
  author    = {Garg, Sanjam and Goel, Aarushi and Jha, Somesh and Mahloujifar, Saeed and Mahmoody, Mohammad and Policharla, Guru-Vamsi and Wang, Mingyuan},
  title     = {Experimenting with Zero-Knowledge Proofs of Training},
  booktitle = {Proceedings of the 2023 ACM SIGSAC Conference on Computer and Communications Security},
  pages     = {1880--1894},
  year      = {2023},
  publisher = {Association for Computing Machinery},
  doi       = {10.1145/3576915.3623101},
  url       = {https://doi.org/10.1145/3576915.3623101}
}

@inproceedings{groth10,
  author    = {Groth, Jens},
  title     = {Short Pairing-Based Non-Interactive Zero-Knowledge Arguments},
  booktitle = {Advances in Cryptology--ASIACRYPT 2010},
  series    = {Lecture Notes in Computer Science},
  volume    = {6477},
  pages     = {321--340},
  year      = {2010},
  publisher = {Springer},
  doi       = {10.1007/978-3-642-17373-8_19},
  url       = {https://doi.org/10.1007/978-3-642-17373-8_19}
}

@article{goldreich1991proofs,
  author    = {Goldreich, Oded and Micali, Silvio and Wigderson, Avi},
  title     = {Proofs that Yield Nothing But Their Validity or All Languages in {NP} Have Zero-Knowledge Proof Systems},
  journal   = {Journal of the ACM},
  volume    = {38},
  number    = {3},
  pages     = {691--729},
  year      = {1991},
  publisher = {Association for Computing Machinery},
  doi       = {10.1145/116825.116852},
  url       = {https://doi.org/10.1145/116825.116852}
}

@misc{zkprooforg,
  author       = {{ZKProof Standards}},
  title        = {{ZKProof.org} -- Open Industry Academic Initiative for Zero-Knowledge Proof Standardization},
  howpublished = {\url{https://zkproof.org}},
  year         = {2024},
  note         = {Accessed: 2026-04-01}
}

@inproceedings{ben1990everything,
  author    = {Ben-Or, Michael and Goldreich, Oded and Goldwasser, Shafi and H{\aa}stad, Johan and Kilian, Joe and Micali, Silvio and Rogaway, Phillip},
  title     = {Everything Provable is Provable in Zero-Knowledge},
  booktitle = {Advances in Cryptology--CRYPTO'88},
  series    = {Lecture Notes in Computer Science},
  volume    = {403},
  pages     = {37--56},
  year      = {1990},
  publisher = {Springer},
  doi       = {10.1007/0-387-34799-2_4},
  url       = {https://doi.org/10.1007/0-387-34799-2_4}
}

@inproceedings{kilian1992note,
  author    = {Kilian, Joe},
  title     = {A Note on Efficient Zero-Knowledge Proofs and Arguments},
  booktitle = {Proceedings of the Twenty-Fourth Annual ACM Symposium on Theory of Computing},
  pages     = {723--732},
  year      = {1992},
  publisher = {Association for Computing Machinery},
  doi       = {10.1145/129712.129782},
  url       = {https://doi.org/10.1145/129712.129782}
}

@article{goldwasser2015delegating,
  author    = {Goldwasser, Shafi and Kalai, Yael Tauman and Rothblum, Guy N.},
  title     = {Delegating Computation: Interactive Proofs for Muggles},
  journal   = {Journal of the ACM},
  volume    = {62},
  number    = {4},
  articleno = {27},
  numpages  = {64},
  year      = {2015},
  publisher = {Association for Computing Machinery},
  doi       = {10.1145/2699436},
  url       = {https://doi.org/10.1145/2699436}
}

@inproceedings{GGPR13,
  author    = {Gennaro, Rosario and Gentry, Craig and Parno, Bryan and Raykova, Mariana},
  title     = {Quadratic Span Programs and Succinct {NIZKs} without {PCPs}},
  booktitle = {Advances in Cryptology--EUROCRYPT 2013},
  series    = {Lecture Notes in Computer Science},
  volume    = {7881},
  pages     = {626--645},
  year      = {2013},
  publisher = {Springer},
  doi       = {10.1007/978-3-642-38348-9_37},
  url       = {https://doi.org/10.1007/978-3-642-38348-9_37}
}

@inproceedings{fiege1987zero,
  author    = {Feige, Uriel and Fiat, Amos and Shamir, Adi},
  title     = {Zero Knowledge Proofs of Identity},
  booktitle = {Proceedings of the Nineteenth Annual ACM Symposium on Theory of Computing},
  pages     = {210--217},
  year      = {1987},
  publisher = {Association for Computing Machinery},
  doi       = {10.1145/28395.28419},
  url       = {https://doi.org/10.1145/28395.28419}
}

@misc{zk-creds,
  author       = {Rosenberg, Michael and White, Jacob and Garman, Christina and Miers, Ian},
  title        = {{zk-creds}: Flexible Anonymous Credentials from {zkSNARKs} and Existing Identity Infrastructure},
  howpublished = {{IACR} Cryptology ePrint Archive, Report 2022/878},
  year         = {2022},
  doi          = {10.1109/SP46215.2023.00108},
  url          = {https://eprint.iacr.org/2022/878}
}

@misc{settyThalerWahbyCCS23,
  author       = {Setty, Srinath and Thaler, Justin and Wahby, Riad},
  title        = {Customizable Constraint Systems for Succinct Arguments},
  howpublished = {{IACR} Cryptology ePrint Archive, Report 2023/552},
  year         = {2023},
  url          = {https://eprint.iacr.org/2023/552}
}

@misc{lurk,
  author       = {Amin, Nada and Burnham, John and Garillot, Fran{\c{c}}ois and Gennaro, Rosario and K{\"u}nzang, Chhi'm\`ed and Rogozin, Daniel and Wong, Cameron},
  title        = {{LURK}: Lambda, the Ultimate Recursive Knowledge},
  howpublished = {{IACR} Cryptology ePrint Archive, Report 2023/369},
  year         = {2023},
  url          = {https://eprint.iacr.org/2023/369}
}

@misc{leo,
  author       = {Chin, Collin and Wu, Howard and Chu, Raymond and Coglio, Alessandro and McCarthy, Eric and Smith, Eric},
  title        = {Leo: A Programming Language for Formally Verified, Zero-Knowledge Applications},
  howpublished = {{IACR} Cryptology ePrint Archive, Report 2021/651},
  year         = {2021},
  url          = {https://eprint.iacr.org/2021/651}
}

@misc{ethSTARK,
  author       = {{StarkWare}},
  title        = {{ethSTARK} Documentation},
  howpublished = {{IACR} Cryptology ePrint Archive, Report 2021/582},
  year         = {2021},
  url          = {https://eprint.iacr.org/2021/582}
}

@misc{nova,
  author       = {Kothapalli, Abhiram and Setty, Srinath and Tzialla, Ioanna},
  title        = {Nova: Recursive Zero-Knowledge Arguments from Folding Schemes},
  howpublished = {{IACR} Cryptology ePrint Archive, Report 2021/370},
  year         = {2021},
  url          = {https://eprint.iacr.org/2021/370}
}

@misc{tinyram,
  author       = {Ben-Sasson, Eli and Chiesa, Alessandro and Genkin, Daniel and Tromer, Eran and Virza, Madars},
  title        = {{SNARKs} for {C}: Verifying Program Executions Succinctly and in Zero Knowledge},
  howpublished = {{IACR} Cryptology ePrint Archive, Report 2013/507},
  year         = {2013},
  url          = {https://eprint.iacr.org/2013/507}
}

@misc{jolt,
  author       = {Arun, Arasu and Setty, Srinath and Thaler, Justin},
  title        = {Jolt: {SNARKs} for Virtual Machines via Lookups},
  howpublished = {{IACR} Cryptology ePrint Archive, Report 2023/1217},
  year         = {2023},
  url          = {https://eprint.iacr.org/2023/1217}
}

@misc{circ,
  author       = {Ozdemir, Alex and Brown, Fraser and Wahby, Riad S.},
  title        = {{CirC}: Compiler Infrastructure for Proof Systems, Software Verification, and More},
  howpublished = {{IACR} Cryptology ePrint Archive, Report 2020/1586},
  year         = {2020},
  url          = {https://eprint.iacr.org/2020/1586}
}

@misc{BitanskySNARK11,
  author       = {Bitansky, Nir and Canetti, Ran and Chiesa, Alessandro and Tromer, Eran},
  title        = {From Extractable Collision Resistance to Succinct Non-Interactive Arguments of Knowledge, and Back Again},
  howpublished = {{IACR} Cryptology ePrint Archive, Report 2011/443},
  year         = {2011},
  url          = {https://eprint.iacr.org/2011/443}
}

@misc{efZKR,
  author = {{Ethereum Foundation}},
  title  = {Zero-Knowledge Rollups},
  year   = {2023},
  url    = {https://ethereum.org/en/developers/docs/scaling/zk-rollups/},
}

@misc{zkSync,
  author = {{Matter Labs}},
  title  = {{zkSync} Era Docs},
  year   = {2023},
  url    = {https://era.zksync.io/docs/reference/},
}

@misc{veedo2023,
  author = {{StarkWare}},
  title  = {{VeeDo}: A {STARK}-Based {VDF} Service},
  year   = {2020},
  url    = {https://medium.com/starkware/presenting-veedo-e4bbff77c7ae},
}

@misc{polygon,
  author = {{Polygon Labs}},
  title  = {{zkEVM} Wiki},
  year   = {2023},
  url    = {https://wiki.polygon.technology/docs/zkevm/},
}

@misc{polygon2,
  author = {{Polygon Labs}},
  title  = {{CDK} Wiki},
  year   = {2023},
  url    = {https://wiki.polygon.technology/docs/cdk/},
}

@misc{starknetCairo,
  author = {{Starknet}},
  title  = {Cairo and Sierra},
  year   = {2023},
  url    = {https://docs.starknet.io/documentation/architecture_and_concepts/Smart_Contracts/cairo-and-sierra/},
}

@misc{starknetAA,
  author = {{StarkWare}},
  title  = {Native Account Abstraction: Opening Blockchain to New Possibilities},
  year   = {2023},
  url    = {https://starkware.co/resource/native-account-abstraction-opening-blockchain-to-new-possibilities/},
}

@misc{starkEx,
  author = {{StarkWare}},
  title  = {StarkEx Documentation: Overview},
  year   = {2023},
  url    = {https://docs.starkware.co/starkex/overview.html},
}

@misc{scroll,
  author = {{Scroll}},
  title  = {Ethereum and Scroll Differences},
  year   = {2023},
  url    = {https://docs.scroll.io/en/developers/ethereum-and-scroll-differences/#evm-opcodes},
}

@misc{linea,
  author = {{Linea}},
  title  = {Linea Documentation},
  year   = {2023},
  url    = {https://docs.linea.build/overview},
}

@misc{telepathy,
  author = {{Telepathy}},
  title  = {Telepathy Documentation},
  year   = {2024},
  url    = {https://docs.telepathy.xyz/},
}

@misc{penumbra,
  author = {{Penumbra}},
  title  = {Penumbra Protocol Documentation},
  year   = {2024},
  url    = {https://protocol.penumbra.zone/main/penumbra.html},
}

@misc{noirLang,
  author = {{Noir}},
  title  = {Noir Language},
  year   = {2023},
  url    = {https://noir-lang.org/},
  note   = {Accessed: November 24, 2023}
}

@misc{aztecDocs,
  author = {{Aztec}},
  title  = {Aztec Documentation},
  year   = {2023},
  url    = {https://docs.aztec.network/},
  note   = {Accessed: November 24, 2023}
}

@misc{herodotus,
  author = {{Herodotus Dev Ltd.}},
  title  = {Herodotus: Secure On-Chain Data},
  year   = {2023},
  url    = {https://herodotus.dev},
}

@misc{filecoin,
  author = {{Protocol Labs Inc.}},
  title  = {{zk-SNARKs} for the World},
  year   = {2021},
  url    = {https://research.protocol.ai/sites/snarks/},
}

@misc{ezkl,
  author = {{Zkonduit Inc.}},
  title  = {What is {EZKL}?},
  year   = {2023},
  url    = {https://docs.ezkl.xyz/},
}

@misc{modulus,
  author = {{Modulus Labs}},
  title  = {Bring Powerful {AI} On-Chain with Specialized {ZK}},
  year   = {2023},
  url    = {https://www.modulus.xyz/},
  note   = {Accessed: December 13, 2023}
}

@misc{buterin2023,
  author       = {Buterin, Vitalik and Illum, Jacob and Nadler, Matthias and Sch{\"a}r, Fabian and Soleimani, Ameen},
  title        = {Blockchain Privacy and Regulatory Compliance: Towards a Practical Equilibrium},
  howpublished = {{SSRN} 4563364},
  year         = {2023},
  url          = {https://ssrn.com/abstract=4563364}
}

@misc{tcash,
  author = {Pertsev, Alexey and Semenov, Roman and Storm, Roman},
  title  = {Tornado Cash Privacy Solution},
  year   = {2019},
  url    = {https://berkeley-defi.github.io/assets/material/Tornado%20Cash%20Whitepaper.pdf},
  note   = {Accessed: December 14, 2023}
}

@misc{pdao,
  author = {Dunaif, Griffin and Boneh, Dan},
  title  = {How to Build a Private {DAO} on Ethereum},
  year   = {2021},
  url    = {https://hackmd.io/nCASdhqVQNWwMhpTmKpnKQ},
  note   = {Accessed: December 14, 2023}
}

@misc{zkapp,
  author = {{Mina Foundation}},
  title  = {How {zkApps} Work},
  year   = {2023},
  url    = {https://docs.minaprotocol.com/zkapps/how-zkapps-work},
  note   = {Accessed: December 14, 2023}
}

@misc{ZCash,
  author = {{Zcash}},
  title  = {Zcash: Privacy-Protecting Digital Currency},
  year   = {2023},
  url    = {https://z.cash/},
  note   = {Accessed: November 24, 2023}
}

@misc{Aleo,
  author = {{Aleo}},
  title  = {Aleo: A New Platform for Private Applications},
  year   = {2023},
  url    = {https://www.aleo.org/},
  note   = {Accessed: November 24, 2023}
}

@misc{Mina,
  author = {{Mina Protocol}},
  title  = {Mina Protocol: The World's Lightest Blockchain},
  year   = {2023},
  url    = {https://minaprotocol.com/},
  note   = {Accessed: November 24, 2023}
}

@misc{zkp2p,
  author = {{ZKP2P}},
  title  = {{ZKP2P}: Exploring Zero-Knowledge Proofs and Peer-to-Peer Technologies},
  year   = {2023},
  url    = {https://zkp2p.xyz/},
  note   = {Accessed: January 23, 2024}
}

@misc{zpass,
  author = {{Aleo}},
  title  = {zPass},
  year   = {2024},
  url    = {https://zpass.docs.aleo.org/zpass/overview},
  note   = {Accessed: January 24, 2024}
}

@misc{galxe,
  author = {{Galxe}},
  title  = {Galxe Protocol Whitepaper},
  year   = {2023},
  url    = {https://github.com/Galxe/protocol-whitepaper},
  note   = {Accessed: January 24, 2024}
}

@misc{worldcoin,
  author = {{Worldcoin}},
  title  = {Intro to Zero-Knowledge Proofs, Semaphore and Their Application in World {ID}},
  year   = {2023},
  url    = {https://worldcoin.org/blog/worldcoin/intro-zero-knowledge-proofs-semaphore-application-world-id},
  note   = {Accessed: January 24, 2024}
}

@misc{proven,
  author = {{Proven}},
  title  = {Cryptographically Proving Financial Health},
  year   = {2023},
  url    = {https://www.proven.tools/products},
  note   = {Accessed: January 30, 2024}
}

@misc{QEDIT,
  author = {{QEDIT}},
  title  = {Zero-Knowledge Proofs},
  year   = {2023},
  url    = {https://qed-it.com/},
  note   = {Accessed: February 1, 2024}
}

@misc{Giza,
  author = {{GIZATECH, INC.}},
  title  = {Actionable {AI} for Decentralized Applications},
  year   = {2024},
  url    = {https://www.gizatech.xyz/},
  note   = {Accessed: April 3, 2024}
}

@misc{circom,
  author = {{Circom}},
  title  = {Circom 2 Documentation},
  year   = {2024},
  url    = {https://docs.circom.io/},
  note   = {Accessed: 2023-12-05}
}

@misc{polygonmiden,
  author = {{Polygon}},
  title  = {Polygon Miden Documentation},
  year   = {2024},
  url    = {https://0xpolygonmiden.github.io/miden-base/introduction.html?search=setup},
  note   = {Accessed: 2024-01-05}
}

@misc{risczero,
  author = {Bruestle, Jeremy and Gafni, Paul and {RISC Zero Team}},
  title  = {{RISC} Zero Whitepaper},
  year   = {2024},
  url    = {https://dev.risczero.com/proof-system-in-detail.pdf},
  note   = {Accessed: 2024-02-22}
}

@misc{ScalingIntelligence,
  author       = {{Modulus Labs}},
  title        = {Scaling Intelligence: Verifiable Decision Forest Inference with Remainder},
  howpublished = {Technical report},
  year         = {2024},
  month        = feb,
  url          = {https://github.com/Modulus-Labs/Papers/blob/master/remainder-paper.pdf}
}

@misc{darkforest,
  author = {{0xPARC}},
  title  = {Dark Forest},
  year   = {2020},
  url    = {https://zkga.me/},
  note   = {Accessed: February 22, 2024}
}

@misc{circomlib,
  author = {{iden3}},
  title  = {circomlib},
  year   = {2018},
  url    = {https://github.com/iden3/circomlib},
  note   = {Accessed: July 22, 2024}
}

@misc{o1js,
  author = {{Mina}},
  title  = {O1JS},
  year   = {2021},
  url    = {https://docs.minaprotocol.com/zkapps/o1js},
  note   = {Accessed: July 22, 2024}
}

@misc{plonky2,
  author = {{Polygon Zero Team}},
  title  = {Plonky2: Fast Recursive Arguments with {PLONK} and {FRI}},
  year   = {2022},
  url    = {https://github.com/0xPolygonZero/plonky2/blob/main/plonky2/plonky2.pdf},
  note   = {Accessed: July 22, 2024}
}

@misc{zkbench,
  author = {Moore, Calum},
  title  = {{ZK} Bench},
  year   = {2023},
  url    = {https://zkbench.dev/},
  note   = {Accessed: July 22, 2024}
}

@misc{zkbenchpaper,
  author       = {Ernstberger, Jens and Chaliasos, Stefanos and Kadianakis, George and Steinhorst, Sebastian and Jovanovic, Philipp and Gervais, Arthur and Livshits, Benjamin and Orr{\`u}, Michele},
  title        = {{zk-Bench}: A Toolset for Comparative Evaluation and Performance Benchmarking of {SNARKs}},
  howpublished = {{IACR} Cryptology ePrint Archive, Report 2023/1503},
  year         = {2023},
  url          = {https://eprint.iacr.org/2023/1503}
}

@misc{arkworks,
  author = {{arkworks contributors}},
  title  = {arkworks {zkSNARK} Ecosystem},
  year   = {2022},
  url    = {https://arkworks.rs}
}

@misc{gnark-v0.9.0,
  author    = {Botrel, Gautam and Piellard, Thomas and El Housni, Youssef and Kubjas, Ivo and Tabaie, Arya},
  title     = {ConsenSys/gnark: v0.9.0},
  year      = {2023},
  month     = feb,
  publisher = {Zenodo},
  doi       = {10.5281/zenodo.5819104},
  url       = {https://doi.org/10.5281/zenodo.5819104},
  note      = {Version: v0.9.0}
}

@misc{halo2zcash,
  author = {{Electric Coin Company}},
  title  = {Zcash Halo2 Book},
  year   = {2020},
  url    = {https://zcash.github.io/halo2/index.html},
}

@misc{zinc,
  author = {{Matter Labs}},
  title  = {Zinc},
  year   = {2020},
  url    = {https://blog.matter-labs.io/release-of-zinc-v0-1-8d949aa9a2f2},
}

@misc{triton,
  author = {{Neptune}},
  title  = {Announcing Triton {VM}},
  year   = {2022},
  url    = {https://neptune.cash/blog/announcing-tvm/},
}

@misc{olavm,
  author = {Sin7Y},
  title  = {Ola{VM}},
  year   = {2022},
  url    = {https://olavm.org/},
}

@misc{powdr,
  author = {{Powdr}},
  title  = {Powdr {zkVM}},
  year   = {2023},
  url    = {https://www.powdr.org/},
}

@misc{valida,
  author = {{Lita}},
  title  = {Lita {zkVM}: Valida},
  year   = {2023},
  url    = {https://www.lita.foundation/infrastructure#valida}
}

@misc{sp1,
  author = {{Succinct Labs}},
  title  = {{SP1} {zkVM}},
  year   = {2024},
  url    = {https://blog.succinct.xyz/introducing-sp1/}
}

@misc{revisitingzkhardware,
  author = {Shlomovits, Omer},
  title  = {Revisiting Paradigm Hardware Acceleration for Zero Knowledge Proofs},
  year   = {2023},
  url    = {https://medium.com/@omershlomovits/revisiting-paradigm-hardware-acceleration-for-zero-knowledge-proofs-16f717a49555}
}

@misc{ingonyama,
  author = {{Ingonyama}},
  title  = {Ingonyama},
  year   = {2022},
  url    = {https://www.ingonyama.com/}
}

@misc{cysic,
  author = {{Cysic}},
  title  = {Cysic},
  year   = {2022},
  url    = {https://cysic.xyz/}
}

@misc{fabric,
  author = {{Fabric}},
  title  = {Fabric Cryptography},
  year   = {2022},
  url    = {https://www.fabriccryptography.com/}
}

@misc{Irreducible,
  author = {{Irreducible}},
  title  = {Irreducible},
  year   = {2022},
  url    = {https://www.irreducible.com/}
}

@misc{nexus,
  author = {Marin, Daniel and Abdalla, Michel and Govereau, Paul and Groth, Jens and Judson, Samuel and Sosnin, Kristian and Vamsi, Guru},
  title  = {Nexus 1.0: Enabling Verifiable Computation},
  year   = {2024},
  url    = {https://nexus.xyz/}
}

@misc{Supranational,
  author = {{Supranational}},
  title  = {Supranational},
  year   = {2023},
  url    = {https://www.supranational.net/}
}

@misc{paradigmzkhardware,
  author = {Konstantopoulos, Georgios},
  title  = {Hardware Acceleration for Zero Knowledge Proofs},
  year   = {2022},
  url    = {https://www.paradigm.xyz/2022/04/zk-hardware}
}

@inproceedings{wust2022platypus,
    author    = {W\"{u}st, Karl and Kostiainen, Kari and Delius, Noah and Capkun, Srdjan},
    title     = {Platypus: A Central Bank Digital Currency with Unlinkable Transactions and Privacy-Preserving
  Regulation},
    booktitle = {Proceedings of the 2022 ACM SIGSAC Conference on Computer and Communications Security},
    series    = {CCS '22},
    year      = {2022},
    pages     = {2947--2960},
    publisher = {ACM},
    doi       = {10.1145/3548606.3560617}
}

@inproceedings{garman2016accountable,
    author    = {Garman, Christina and Green, Matthew and Miers, Ian},
    title     = {Accountable Privacy for Decentralized Anonymous Payments},
    booktitle = {Financial Cryptography and Data Security (FC 2016)},
    series    = {Lecture Notes in Computer Science},
    volume    = {9603},
    pages     = {81--98},
    year      = {2017},
    publisher = {Springer},
    doi       = {10.1007/978-3-662-54970-4_5}
}

@inproceedings{narula2018zkledger,
    author    = {Narula, Neha and Vasquez, Willy and Virza, Madars},
    title     = {{zkLedger}: Privacy-Preserving Auditing for Distributed Ledgers},
    booktitle = {15th {USENIX} Symposium on Networked Systems Design and
                 Implementation ({NSDI} '18)},
    year      = {2018},
    pages     = {65--80},
    publisher = {{USENIX} Association},
    address   = {Renton, WA},
    isbn      = {978-1-939133-01-4},
    url       = {https://www.usenix.org/conference/nsdi18/presentation/narula}
}

\end{document}